\documentclass[paper]{JHEP3}
\pdfoutput=1
\usepackage{amsmath,amssymb,amsthm,amscd,graphicx}
\input epsf.sty

\theoremstyle{definition}

\newcommand{\CC}{{\cal C}}

\newcommand{\CF}{{\cal F}}

\newcommand{\CH}{{\cal H}}

\newcommand{\CL}{{\cal L}}

\newcommand{\CN}{{\cal N}}
\newcommand{\CO}{{\cal O}}

\newcommand{\CU}{{\cal U}}

\newcommand{\CW}{{\cal W}}

\def\IZ{{\mathbb Z}}
\def\IR{{\mathbb R}}
\def\IC{{\mathbb C}}
\def\IP{{\mathbb P}}
\def\IT{{\mathbb T}}
\def\IS{{\mathbb S}}
\def\IF{{\mathbb F}}

\newcommand{\re}{{\rm e}}
\newcommand{\ri}{{\rm i}}
\newcommand{\rd}{{\rm d}}
\renewcommand{\d}{\partial}

\newcommand{\be}{\begin{equation}}
\newcommand{\ee}{\end{equation}}
\newcommand{\ba}{\begin{aligned}}
\newcommand{\ea}{\end{aligned}}
\newcommand{\ben}{\begin{eqnarray}\displaystyle}
\newcommand{\een}{\end{eqnarray}}

\newcommand{\sectiono}[1]{\section{#1}\setcounter{equation}{0}}

\newdimen\tableauside\tableauside=1.0ex
\newdimen\tableaurule\tableaurule=0.4pt
\newdimen\tableaustep
\def\phantomhrule#1{\hbox{\vbox to0pt{\hrule height\tableaurule width#1\vss}}}
\def\phantomvrule#1{\vbox{\hbox to0pt{\vrule width\tableaurule height#1\hss}}}
\def\sqr{\vbox{%
  \phantomhrule\tableaustep
  \hbox{\phantomvrule\tableaustep\kern\tableaustep\phantomvrule\tableaustep}%
  \hbox{\vbox{\phantomhrule\tableauside}\kern-\tableaurule}}}
\def\squares#1{\hbox{\count0=#1\noindent\loop\sqr
  \advance\count0 by-1 \ifnum\count0>0\repeat}}
\def\tableau#1{\vcenter{\offinterlineskip
  \tableaustep=\tableauside\advance\tableaustep by-\tableaurule
  \kern\normallineskip\hbox
    {\kern\normallineskip\vbox
      {\gettableau#1 0 }%
     \kern\normallineskip\kern\tableaurule}%
  \kern\normallineskip\kern\tableaurule}}
\def\gettableau#1{\ifnum#1=0\let\next=\null\else
\squares{#1}\let\next=\gettableau\fi\next}

\newcommand{\figref}[1]{Fig.~\protect\ref{#1}}

\title{Large $N$ duality beyond the genus expansion}

\author{
Marcos Mari\~no$^{a,b}$, Sara Pasquetti$^c$ and Pavel Putrov$^b$
\\
$^a$ D\'epartement de Physique Th\'eorique et $^b$Section de Math\'ematiques,\\
Universit\'e de Gen\`eve, Gen\`eve, CH-1211 Switzerland\\
\\
$^c$Departement of Physics, CERN,\\
Gen\`eve, CH-1211 Switzerland\\
\\
\email{marcos.marino@unige.ch}, \quad
\email{sara.pasquetti@cern.ch}, \quad
\email{pavel.putrov@unige.ch}
}

\abstract{
We study non-perturbative aspects of the large $N$ duality between Chern--Simons theory and topological strings, and we find 
a rich structure of large $N$ phase transitions in the complex plane of the 't Hooft parameter. These transitions are due to large $N$ 
instanton effects, and they can be regarded as a deformation of the Stokes phenomenon. Moreover, we show that, for generic values of the 't Hooft coupling, instanton effects are not exponentially suppressed at large $N$ 
and they correct the genus expansion. This phenomenon was first discovered in the context of matrix models, and we interpret it as a generalization of the oscillatory asymptotics 
along anti--Stokes lines. In the string dual, the instanton effects can be interpreted as corrections to the saddle string geometry due to discretized neighbouring geometries. 
As a mathematical application, we obtain the $1/N$ asymptotics of the partition function of Chern--Simons 
theory on the lens space $L(2,1)$, and we test it numerically to high precision in order to exhibit the importance of instanton effects.
}    

\begin{document}

\sectiono{Introduction}

According to large $N$ dualities, the $1/N$ expansion of different observables in $U(N)$ gauge theories can be reinterpreted as a genus expansion in an appropriate string 
theory. For example, the gauge theory free energy  
at finite volume has an expansion of the form 
\be
F(g_s, N)\sim \sum_{g=0}^{\infty} g_s^{2g-2} F_g(t), 
\ee
where $g_s$ is the gauge theory coupling constant and $t=g_s N$ is the 't Hooft parameter. In large $N$ dualities, $g_s$ is identified with the string coupling constant, the 
't Hooft parameter becomes a geometric modulus of the string target space, and the amplitudes $F_g(t)$ are identified with free energies at genus $g$ of a string theory. 

The $1/N$ expansion in gauge theory is an asymptotic expansion, and it is expected to have corrections of the form 
\be
\label{correctedf}
F(g_s, N)\sim \sum_{g=0}^{\infty} g_s^{2g-2} F_g(t)+\CO(\re^{-A(t)/g_s}).
\ee
These corrections are invisible in the traditional $1/N$ asymptotics\footnote{By traditional or classical asymptotics, we mean asymptotic expansions in which non-analytic terms are not taken into account. There are refinements of traditional asymptotics, which go sometimes under the name of ``hyperasymptotics," where these terms are included in a 
systematic way, see \cite{hyper} for an overview.}, and they are associated to large $N$ instantons. Although large $N$ instantons are typically built upon classical gauge theory instantons, they should not be confused with them. A classical instanton is a saddle point of the classical action, while a large $N$ instanton is a saddle point of the large $N$ 
effective action; see \cite{affleck,munster} and \cite{gm} for explicit examples in the $\IC\IP^N$ model and in two-dimensional Yang--Mills theory, respectively. 
In particular, the action $A(t)$ of a large $N$ instanton is a non-trivial function of the 't Hooft parameter which includes quantum corrections at all loops. 
In some cases, large $N$ instantons can be interpreted 
in terms of D-branes of the string theory dual, as it happens for example in matrix models of non-critical strings. 

One could think that the instanton corrections in (\ref{correctedf}) are unimportant since they are exponentially suppressed at large $N$, 
but this is not always the case. As we move in parameter 
space they might become of order one and start contributing to the large $N$ asymptotics. This scenario 
was advocated long ago by Neuberger \cite{neuberger} in order to explain the occurrence of critical points in Hermitian matrix models and 
the third-order large $N$ phase transitions 
of unitary matrix models \cite{gw,wadia}. The same mechanism explains as well the Douglas--Kazakov transition in two-dimensional Yang--Mills on the sphere \cite{dk,gm}. In these instanton-driven phase transitions, as we change 
the 't Hooft parameter at fixed $g_s$, the action of the instanton $A(t)$ vanish at some finite value $t=t_c$. This leads in general to different $1/N$ expansions
for the regions $t>t_c$ and $t<t_c$. 

On the other hand, if we regard $g_s$ as a complex variable, we should expect that the $1/N$ asymptotics changes discontinuously 
as we change the argument of $g_s$ (and keep $|t|$ fixed). In classical asymptotic analysis, this discontinuous change is the well-known Stokes phenomenon.   
The reason for this phenomenon is in fact the same one that underlies instanton-induced 
phase transitions: an exponentially small quantity, due to a sub-leading saddle point, becomes less and less suppressed as we change the argument of the 
expansion variable. Along the so-called anti-Stokes line, the contribution of this saddle is of order one and has to be included in the asymptotics.

In general, a large $N$ gauge theory with instanton sectors, 
should display a complex pattern of large $N$ phase transitions, combining the instanton-driven phase transitions at fixed $g_s$ with 
the Stokes-like transitions as we vary the argument of $g_s$. If the gauge theory that we are studying has a string dual in at least one of the phases, it is certainly interesting to 
understand what are the implications of these transitions in the string picture. 

In this paper we will analyze these issues in a family of models which have well-understood large $N$ duals, namely Chern--Simons theory on the lens space $L(p,1)$. 
This gauge theory is described by topological string theory on a Calabi--Yau space given by an $A_p$ fibration over a two-sphere \cite{gv,akmv,hy,hoy}. 
However, this large $N$ string dual is intrinsically semiclassical, since it 
describes a generic but fixed saddle--point of the gauge theory. A full non-perturbative study of this duality, where we sum over all saddle-points, reveals a surprisingly rich phase structure as we move in the complex plane of the 't Hooft coupling. The main results of our analysis are the following: 

\begin{enumerate}

\item For each complex value of $t$, the large $N$ asymptotics of the non-perturbative free energy is dominated by a fixed string target geometry, i.e. by a point in the Calabi--Yau moduli space. However, this ``saddle geometry" might change as we vary $t$. 
 
\item For some values of $t$, the $1/N$ asymptotics is given by a conventional genus expansion. There are non-perturbative effects due to the contributions of 
neighbouring geometries, but they are exponentially suppressed. For other values of $t$, the asymptotics has an oscillatory behavior due to large $N$ instantons. This type of behavior was first observed in the context of matrix models in \cite{bde}, and further studied in \cite{eynard,em}. In particular, the large $N$ asymptotics is no longer given by a genus expansion around a fixed geometry, and corrections due to neighboring geometries are crucial, already at the next-to-leading order. 

\item The change of saddle geometry as we change the complex 't Hooft parameter can be regarded as a generalized or ``deformed" Stokes phenomenon. 
In the limit of vanishing 't Hooft coupling, 
the asymptotics changes discontinuously as we change the argument of the string coupling constant, and we recover the classical Stokes phenomenon. For finite $t$ the phenomenon is smoothed out, and the asymptotics changes continuously. 

\end{enumerate}

In practice we have focused on the simplest, nontrivial model in the family, namely $L(2,1)=\IR\IP^3$. The gauge theory saddle points are characterized by a symmetry breaking pattern $U(N) \rightarrow U(N_1) \times U(N_2)$. The dual Calabi--Yau manifold is the so-called local $\IF_0$ geometry, where $\IF_0=\IP^1\times \IP^1 $ is a Hirzebruch surface. 
It has two K\"ahler parameters, corresponding at large radius to the sizes of the two $\IP^1$s in $\IF_0$. The partial 't Hooft couplings $t_1$, $t_2$, where $t_i =g_s N_i$, parametrize the K\"ahler moduli space near the orbifold point described in \cite{akmv}, and their sum $t_1+t_2$ is fixed to be the total 't Hooft parameter of the gauge theory. Each saddle geometry is characterized by an ``equilibrium value" for $t_2$. The phase diagram in the region $|t|>0$, $0\le {\rm Arg}(g_s)\le \pi/2$ is shown in 
\figref{phasediagram}. 

Our analysis shows very clearly that the large $N$ instantons analyzed in \cite{bde, mswone,mswtwo, mmnp,eynard, em,ps} are crucial in order to understand 
large $N$ dualities. To be precise, for generic values of the `t Hooft parameter, the correct large $N$ asymptotics of the exact gauge theory partition function 
involves these instanton effects already at next-to-leading order. In order to dispel any doubt about this, 
we study in detail the partition function of Chern--Simons theory on $\IR\IP^3$, for imaginary values of the coupling constant. These are precisely the the ``physical" or ``on-shell" values of the coupling constant in Chern--Simons theory. Our analysis shows that these values correspond to an anti-Stokes line with oscillatory asymptotics, in which the traditional genus expansion 
has to be supplemented with instanton corrections. We test numerically our predictions for the instanton corrected asymptotics against the 
exact non-perturbative Chern--Simons partition function, and we obtain an impressive agreement. 

Our analysis also clarifies the issue of background dependence in topological string theory. Let us focus again on the example of Chern--Simons theory on 
$\IR \IP^3$. Perturbative topological string theory depends on a choice of background $(t_1, t_2)$. The gauge theory partition function is background independent, 
in the sense that it only depends 
on the total 't Hooft parameter $t=t_1+t_2$ (which is, as needed for consistency, a modular invariant in the dual topological string). 
However, the large $N$ asymptotics of the partition 
function is ``peaked" around a particular value of $(t_1, t_2)$ which depends on $t$ and corresponds to a particular saddle geometry. This fixed background 
emerges as an equilibrium value as a consequence of the large $N$ limit, but it has no meaning at finite $N$. In fact, in the full theory, the value of the background (specified by, 
say, the value of $t_2$) is not fixed, since $t_2$ is rather an internal or fluctuating variable which we have to sum over. But as usual in statistical mechanics, in the thermodynamic limit 
$N\rightarrow \infty$ this fluctuating variable is peaked around an equilibrium value.

We would like to point out that the role of instanton effects in the $1/N$ expansion and their string theory interpretation has been addressed before in different contexts. 
For example, the instanton-induced Douglas--Kazakov phase transition has been interpreted as a breakdown of the large $N$ string dual \cite{dk}. In AdS/CFT at finite temperature, 
an incarnation of the Gross--Witten--Wadia phase transition has been argued to correspond to a breakdown of the geometric description of the string target \cite{alvarezgaume}\footnote{The Hawking--Page transition in AdS/CFT is a large $N$ phase transition leading to a change 
of topology, but it involves effects of order $\exp(-N^2)$ and it is not driven by instantons.}. 
The importance of exponentially small effects, and their effects on the string target space, has been emphasized in \cite{mmss}, which studies as well 
the Stokes phenomenon in the open string moduli space. In \cite{kyriakos}, a large $N$ phase transition in $\CN=4$ super Yang--Mills theory on K3 was found, separating a phase 
dominated by the zero-instanton sector from a phase dominated by an instanton sector with large instanton number. In the context of matrix models, general techniques to analyze large $N$ phase transitions have been developed along the years, culminating in the formulation in terms of Boutroux spectral curves in \cite{bertola}. Of course, the fact that the genus expansion can not capture the large $N$ asymptotics was discovered in the 
context of matrix models and explained in detail in \cite{bde}. 

This paper is organized as follows: 
In Section 2  we review the structure of instantons corrections 
in matrix models and topological strings.
In Section 3 we present our model, the Chern--Simons theory on the lens space
$L(p,1)$, and  its matrix model realization.
We then specialize to the $p=2$  case and review
the large $N$ duality between the perturbative expansion 
of the matrix model in a fixed filling fraction $(N_1,N_2)$
configuration and the  topological string on the local
 $\mathbb{P}^1\times \mathbb{P}^1$ geometry, in a fixed background $t_1,t_2$.
In Section 4 we study  the large $N$ phase diagram.
We start by considering two particular cases: real and imaginary $g_s$,
which correspond to two phases of the gauge theory with two distinct dominant saddle configurations, each dual to a fixed background geometry.
By following the analysis of Section 2, we explain the structure of the 
large $N$ asymptotic of the two phases and provide very precise 
numerical checks of our predictions.
We then complete the study of the phase diagram for generic complex $g_s$
and interpret the large $N$ phase transitions between the various  phases 
as due to the Stokes phenomen.
We end this section by commenting on the issue of the background independence.
In Section 5 we discuss the phase diagram of the cubic matrix model.
Thanks to its close relation with the Airy function,
the prototypical example of the Stokes phenomenon,
this example clarifies the nature of the large $N$
phase transitions between dominant saddles/backgrounds which we found in Chern--Simons theory.
Section 6 contains conclusions and a list of open 
issues.
Finally in the appendix we collect some details of the model for different submanifolds of moduli space.

\sectiono{Instanton corrections in matrix models and topological strings}

In this section we review instanton corrections in matrix models/topological strings, following mainly \cite{bde,mswtwo,em}. 

Multi-cut matrix models and topological strings are characterized, at the perturbative level, by genus $g$ amplitudes of the form 
\be
F_g(t)=F_g(t_1, \cdots, t_s), 
\ee
where, in the case of matrix models, $t_i$ are partial 't Hooft couplings 
\be
t_i=g_s N_i, \quad i=1, \cdots, s,
\ee
 $N_i$ is the number of eigenvalues in the $i$-th cut
 and $s$ is the number of cuts. The perturbative partition function is given, as usual, by 
\be
\label{smallgs}
Z_{\rm p}(N_1, \cdots, N_s)=\exp \biggl\{ \sum_{g=0}^{\infty} F_g(t)g_s^{2g-2} \biggr\}. 
\ee
A choice of moduli or 't Hooft parameters $t=(t_1, \cdots, t_n)$ in this partition function is called a choice of {\it background}. In matrix models and in gauge theory, the quantity 
\be
t=\sum_{i=1}^n t_i =g_s N
\ee
where $N$ is the rank of the $U(N)$ gauge group, is fixed. Background-independent quantities should only depend on $g_s$ and $N$, while 
the perturbative partition function $Z_{\rm p}$ depends on a choice of $t_i$ and is therefore a background dependent quantity. 

The {\it non-perturbative partition function} $Z^{\CC}_{\rm np}$ was introduced in the context of matrix models with a polynomial potential $V(z)$ 
in \cite{david,bde,eynard}, and it depends on a choice of contour $\CC$ in the complex 
plane. This contour is the integration path for the eigenvalues of the matrix integral. We then have
\be
Z^{\CC}_{\rm np}(N, g_s)={1\over N!} \int_{\CC} \prod_{i=1}^N{\rd \lambda_i \over 2\pi}\Delta^2(\lambda) \re^{-{1\over g_s} \sum_{i=1}^N V(\lambda_i)} 
\ee
where $\Delta(\lambda)$ is the Vandermonde determinant. Of course, 
the contour $\CC$ is chosen in such a way that this integral converges. By standard saddle--point techniques, we can always deform the path $\CC$ into a sum of 
paths $\CC_k$ which go through the critical points of $V(z)$ and are paths of steepest descent \cite{fr}, 
\be
\CC=\sum_{k=1}^s \zeta_k \CC_k, 
\ee
where we have assumed that $V(z)$ has $s$ critical points. Therefore
\be
\label{ccp}
Z^{\CC}_{\rm np}(N, g_s) =\sum_{N_1 +\cdots +N_s=N} \zeta_1^{N_1} \cdots \zeta_s^{N_s} Z_{\rm p} (N_1, \ldots, N_s).
\ee
where
\be
\label{pzmi}
Z_{\rm p}(N_1, \ldots, N_s) = {1 \over N_1! \cdots N_s!} \int_{\lambda^{(1)}_{k_1} \in \CC_1} \cdots \int_{\lambda^{(s)}_{k_s} \in \CC_s} \prod_{i=1}^N {\rd\lambda_i \over 2 \pi}\, \Delta^2(\lambda)\, \re^{-{1\over g_s} \sum_{i=1}^N V(\lambda_i)}.
\ee
To write this equation, we have split the $N$ eigenvalues into $s$ sets of $N_I$ eigenvalues, $I=1, \ldots, s$, which are denoted by
\be
\{ \lambda^{(I)}_{k_I} \}_{k_I=1, \dots, N_I}, \qquad I=1, \ldots, s. 
\ee
Each of the integrals (\ref{pzmi}) has a small $g_s$ asymptotic expansion given by (\ref{smallgs}). 

Equation (\ref{ccp}) expresses a background-independent quantity $Z^{\CC}_{\rm np}$ as a sum of background-dependent quantities $Z_{\rm p}(t)$. This sum can be in turn regarded as a sum over matrix model instantons, which have been identified long ago in terms of eigenvalue tunneling \cite{david,shenker}. Formally, we can write the sum (\ref{ccp}) as the perturbative partition function coming from a {\it fixed} background $N_i=N_i^*$, plus an infinite sums of corrections for the remaining values of $N_i$. These corrections 
are non-perturbative in $g_s$. We 
write (\ref{ccp}), schematically, as
\be
\label{ccnext}
Z^{\CC}_{\rm np}(N, g_s) =Z_{\rm p}(N_1^*, \cdots, N_p^*) +\CO(\re^{-1/g_s})
\ee

In order to be more concrete, we will restrict ourselves to models where $s=2$, or 
equivalently, models with two moduli $t_1$, $t_2$. Up to an overall normalization, we can set $\zeta_1=1$, $\zeta_2=\zeta$. 
The detailed form of (\ref{ccnext}) depends on the choice of background. If the background is on the {\it boundary}, we have 
$(N^*_1, N^*_2)=(N,0)$ or $(0,N)$. Otherwise, we say that the background is an {\it interior point}. The expression for (\ref{ccnext}) when the background is on the boundary was worked 
out in \cite{mswtwo}. Assuming for concreteness that the boundary is at $(N,0)$, we have
\be
\label{npboundary}
\ba
Z^{\CC}_{\rm np}(N, g_s) &=Z_{\rm p}(N)\biggl(1 +\sum_{\ell\ge 1} {g_s^{\ell^2/2} \over (2\pi)^{\ell/2}}\, G_2(\ell+1)\, \zeta^\ell\, \hat q^{\frac{\ell^2}{2}}\, \exp \left( - \frac{\ell  A}{g_s} \right) \\
& \cdot \sum_{k} \sum_{m_i>0}\sum_{g_i>1-{m_i\over 2}} {g_s^{\sum_i (2g_i+m_i-2)}\over k! m_1!\,\dots\, m_k!}\,\,\, 
\widehat F_{g_1}^{(m_1)}\dots \widehat F_{g_k}^{(m_k)}  (-\ell)^{\sum_i l_i}\biggr).
\ea
\ee
In this equation we have introduced the following notations. $Z_{\rm p}(N)$ is the partition function for the one-cut model where all eigenvalues sit in the first critical point. 
The functions $\widehat F_g(t_1, t_2)$ are defined by
\be
F_g(t_1, t_2) = F^{\rm G}_g(t_2) + \widehat F_g(t_1, t_2), 
\ee
where $F^{\rm G}_g(t)$ are the genus $g$ free energies of the gauged Gaussian matrix model with 't~Hooft parameter $t$. $\widehat F_{g}^{(m)}$ denotes the $m$-th derivative 
of $F_g$ w.r.t. $s$, which is defined as
\be
\label{average}
s={1\over 2}(t_1 -t_2). 
\ee
All derivatives are evaluated at $t_1=t$ and $t_2=0$. $G_2(\ell+1)$ is the Barnes function
\be
G_2(\ell+1)=\prod_{n=0}^{\ell-1} n!.
\ee
Finally, 
\be\label{hata}
A(t) = \partial_{s}  F_0 \qquad \mathrm{and} \qquad \hat q = \exp \Bigl(  \partial_s^2 \widehat F_0 \Bigr).
\ee
At leading order in $g_s$, we have
\be
\ba
&\sum_{k} \sum_{m_i>0}\sum_{g_i>1-{m_i\over 2}} {g_s^{\sum_i (2g_i+m_i-2)}\over k! m_1!\,\dots\, m_k!}\,\,\, \widehat F_{g_1}^{(m_1)}\dots \widehat F_{g_k}^{(m_k)}  (-\ell)^{\sum_i l_i}
\\
&=1 - g_s \Bigl( \ell\, \partial_{s} \widehat F_1 (t) + {\ell^3 \over 6}\, \partial_s^3 \widehat F_0 (t) \Bigr) + \CO(g_s^2).
\ea
\ee

The expansion around an {\it interior} point was studied in \cite{bde,eynard,em}. It is given by
\be
\label{npinterior}
\ba
& Z^{\CC}_{\rm np}(N, g_s)
= Z_{\rm p}(N_1^*, N_2^*) \sum_{k} \sum_{m_i>0}\sum_{g_i>1-{m_i\over 2}} {g_s^{\sum_i (2g_i+m_i-2)}\over k! m_1!\,\dots\, m_k!}\,\,\, F_{g_1}^{(m_1)}\dots F_{g_k}^{(m_k)}  \,\, \Theta_{\mu,\nu}^{(\sum_i l_i)}(F'_0/g_s,\tau) \\
&\quad =Z_{\rm p}(N_1^*, N_2^*)  \biggl\{ \Theta_{\mu,\nu} +g_s \Bigl(\Theta'_{\mu,\nu} F_1' + {1\over 6} \Theta_{\mu,\nu}'''\,F_0'''\Bigr) + \CO(g_s^2) \biggr\}. 
\ea
\ee
The derivatives of the free energies $F_g$ are again w.r.t. $s$ defined in (\ref{average}). 
The theta function $\Theta_{\mu,\nu}$ with characteristics $(\mu,\nu)$ is defined by
\be
\label{biget}
\Theta_{\mu,\nu}(u,\tau) = \sum_{n\in {\mathbb Z}} {\rm e}^{(n+\mu-N \epsilon)u}\,\,\re^{\pi \ri  (n +\mu-N \epsilon)\tau (n+\mu-N\epsilon)}\,\,{\rm e}^{2 \ri\pi (n+\mu) \nu}
\ee
and it is evaluated at 
\be
u={F_0'(N_i^*)\over g_s},
\qquad
\tau =  {1\over 2\pi \ri} F_0''.
\ee
In the above equation, we have denoted
\be
\epsilon={N_2^* \over N}, \qquad \zeta={\rm e}^{2 \ri\pi\nu}
\ee
in order to make contact with the notations of \cite{em}. In (\ref{npinterior}), the derivatives of the theta function (\ref{biget}) are w.r.t. $u$, therefore 
each derivative introduces a factor of $n+\mu-N\epsilon$ in the sum (\ref{biget}). Notice that, as emphasized in \cite{bde}, the expression (\ref{npinterior}) gives 
a large $N$ asymptotics which goes beyond the genus expansion. In particular, the leading terms in the free energy are of the form, 
\be
F=g_s^{-2} F_0(N_i^*)+ F_1(N_i^*) + \log  \Theta_{\mu,\nu} +\cdots,
\ee
and the theta function leads typically to an oscillatory large $N$ asymptotics. 

The above expressions, (\ref{npboundary}) and (\ref{npinterior}), are for the moment being merely formal. In particular, it is not clear if they provide reasonable asymptotic expansions 
of the original, non-perturbative partition function. There are again two different cases, applying to (\ref{npboundary}) and (\ref{npinterior}):
\begin{enumerate}
\item The expression (\ref{npboundary}) gives an admissible asymptotic expansion if 
\be
\label{stableboundary}
{\rm Re}\, \Bigl( {A(t) \over g_s} \Bigr)>0. 
\ee
In this case, the classical, small $g_s$ asymptotic expansion is given by the genus expansion of $Z_{\rm p}(N)$. The terms with $\ell>1$ give exponentially suppressed 
corrections to the asymptotics. 

\item The expansion around a generic point $N_i^*$ in (\ref{npinterior}) is admissible if the filling fractions, as determined from the spectral curve, are real, i.e.
\be
\label{real}
\epsilon_i={N^*_i \over N}  \in \IR,
\ee
and if 
\be
\label{realvanish}
{\rm Re} \,  \Bigl({F_0'(N_i^*)\over g_s} \Bigr)=0. 
\ee
\end{enumerate}

The conditions (\ref{real}), (\ref{realvanish}) were first spelled out in detail in \cite{david,davidvacua}. As noticed in \cite{davidvacua}, they can be written as 
\be
\label{boutrouxc}
{\rm Im}\biggl\{  {1\over t}  \oint_{\gamma} {\rd x \over 2\pi \ri } y(x)\biggr\}=0, 
\ee
where $y(x)$ is the spectral curve of the matrix model, and $\gamma$ is any cycle on it. In writing these equations, we have followed the conventions of 
\cite{leshouches}, so that 
\be
y^2(x) =V'(x)^2 +\cdots. 
\ee
 A curve with the property (\ref{boutrouxc}) is called a {\it Boutroux curve}. The condition (\ref{realvanish}) makes sure that the term involving $u$ in the theta function (\ref{biget}) is 
 oscillatory. This in turn guarantees that the sum over $n$ will be peaked around 
 \be
 n+\mu-N \epsilon \approx 0
 \ee
 so that the derivatives of the theta function involved in (\ref{npinterior}) are of order one, see \cite{eo} for a related discussion. The condition that the curve is Boutroux gives a nice geometrization 
 of the search for good asymptotic expansions in matrix model theory, and it has been recently developed in much detail in \cite{bertola}. 

The above conditions imply that, in order to have a well-defined large $N$ asymptotics, the backgroud can not be arbitrary. As we change the parameters of the theory, this ``dominant background" will change. In particular, it can change from a boundary point to an interior point. We will see examples of this in the Chern--Simons model discussed in this paper. 

The search for a dominant background is however subtle, since as it is well-known, the perturbative amplitudes $F_g(t_i)$ have a finite radius of convergence. 
The full space of 't Hooft 
parameters has a nontrivial global structure. As we reach the boundary of 
the convegence region, we have to perform a {\it duality transformation} to a different frame. In particular, the condition (\ref{realvanish}) is attached to a particular frame and region, 
in which $F_0$ is well-defined. One could worry that the search for dominant backgrounds is not a well-posed problem, globally. However, we will now argue that 
this is not the case for two reasons. The first one 
is that the condition on the curve being Boutroux is invariant under duality transformations. Second, as shown in \cite{em}, each term in the $1/N$ expansion of (\ref{npinterior}) is 
invariant under duality transformations, up to a phase. This means, in particular, that 
\be
\Bigl| \re^{g_s^{-2} F_0 + F_1} \Theta_{\mu, \nu} \Bigr| 
\ee
is invariant and it is therefore a well-defined quantity in the global moduli space. Looking for the maxima of this function on the subspace of Boutroux curves in moduli space is therefore 
a well-posed global problem, i.e. independent of the duality frame. In each region of moduli space in which $F_0, F_1, \Theta_{\mu,\nu}$ are not singular, the maxima at large $N$ satisfy 
the condition (\ref{realvanish}). Therefore, the problem of solving (\ref{real}), (\ref{realvanish}) can be lifted globally in a consistent way.

The structure of instanton corrections that we have reviewed here was originally derived in the context of matrix models. However, it was proposed in \cite{mswone,mmnp,em} that, since 
it only depends formally on the data of the spectral curve, these corrections should be present in topological strings on local geometries, including toric geometries. In this sense, a suitable  
example is provided by topological string theory on $A_p$ fibrations on $\IP^1$. These models have a non-perturbative 
definition in terms of $U(N)$ Chern--Simons gauge theory on a lens space \cite{akmv}, and they can be explicitly reformulated as a multi-cut matrix integral. They constitute a privileged 
arena for constructing the large $N$ asymptotics presented above and for studying the role of instanton corrections. In the next section we present what is known about these models from the point of 
view of Chern--Simons theory, matrix models, and topological string theory. 

\section{Large $N$ duality for Chern--Simons theory}

In this paper we will study Chern--Simons theory on the lens space $L(p,1)$.
This model has  an open string description as an A--type topological string defined  on a  Calabi--Yau threefold given by the cotangent bundle
over the lens space $T^*L(p,1)$. This is just a particular example of the more general realization, due to Witten \cite{wittencs}, of  Chern--Simons theory on a three-manifold $M$ as a type  A topological string on $T^*M$.

For $M=\IS^3$, which corresponds to the lens space with $p=1$, 
the Gopakumar and Vafa large $N$ duality \cite{gv} provides a description of $U(N)$ Chern--Simons gauge theory in terms of 
closed topological string theory on the resolved conifold 
$\mathcal{O}(-1)+\mathcal{O}(-1)\to \mathbb{P}^1$. This duality can be generalized \cite{akmv} by performing a $\IZ_p$ 
quotient on both sides of the Gopakumar--Vafa duality. 
This leads to a duality between Chern--Simons theory on $S_3/\mathbb{Z}_p\equiv L(p,1)$ 
and the  closed topological A--model string on an  $A_p$ fibration over  $\mathbb{P}^1$.
This target space, unlike the resolved conifold, has a non trivial 
 K\"ahler moduli space and  besides the large radius phase it has non--geometric phases. In particular, 
 the large $N$ duality relates  the perturbative regime of the Cherns--Simons gauge theory---small 't Hooft couplings---to the orbifold regime of the string theory.
This duality has been tested in \cite{akmv} by comparing the perturbative expansion of Chern--Simons theory around a fixed flat connection, 
to the orbifold topological strings amplitudes computed by mirror symmetry. Further tests of the duality have been done in \cite{hy,hoy}. 
In the following we will review some of these results.
We will however go beyond  the perturbative side of the duality and we will study the exact, non-perturbative gauge theory partition function. 
As we will see, in the full theory, the choice of filling fractions is not arbitrary, but it is 
dictated by the only parameters of the theory, the Chern--Simons level $k$ and the rank of the gauge group $N$.

\subsection{Chern--Simons theory on lens spaces}

The lens space $L(p,1)$ is a three--manifold that can be obtained by gluing two solid 2-tori along their  boundaries after performing the 
${\rm SL}(2,\IZ)$ transformation,
\be
\label{lensmat}
U_p =\begin{pmatrix}1 &  0 \\
p & 1 \end{pmatrix}.
\ee
This description makes it possible to calculate the partition function of Chern--Simons theory on these spaces \cite{witten}.
${\rm SL}(2,\IZ)$ transformations lift to operators acting on $\CH(\IT^2)$, the Hilbert space obtained by canonical quantization of the Chern--Simons theory on the two-torus. This space is the 
space of integrable representations of a Wess--Zumino--Witten (WZW) model with gauge group $G$ at
level $k$, where $G$ and $k$ are respectively the Chern--Simons gauge group and the  quantized coupling constant.
In particular the partition function of the $U(N)$
 Chern--Simons theory on the lens space $L(p,1)$ is given by:
\be
Z(L(p,1)) = \langle0 |\CU_p |0\rangle, 
\ee
where $\CU_p$ is the lift of (\ref{lensmat}) to an operator on $\CH(\IT^2)$ and $|0\rangle$ is the vacuum state, corresponding to the trivial representation. This partition function  has the following explicit form
(up to an overall constant) \cite{rz}:
\be
Z= {(-1)^{|\Delta_+|} \re^{-{2\pi \ri \over \hat k p} \rho^2}\over (-\ri p \hat k)^{N/2}} \sum_{n \in \IZ^N/p \IZ^N}
\sum_{w \in {\cal W}} \epsilon (w) \exp
\Bigl\{ {\ri \pi \over \hat k  p} ( \rho^2 - 2\rho\cdot(\hat k 
 n +w(\rho)) + (\hat k n +w(\rho))^2
)\Bigr\}. 
\ee
In this expression, $\hat k=k +y$, where $y$ denotes the dual Coxter number, $|\Delta_+|$ is the number of positive roots of $
G$, and $\rho$ is the Weyl vector. 
Using Weyl's formula
\be
\sum_{w\in \CW} \epsilon (w) \re^{w(\rho)}=\prod_{\alpha>0} 2\sinh {\alpha\over 2}
\ee
we can also  write the partition function as:
\be
\label{znp}
Z=
{\ri^{|\Delta_+|} \over (-\ri p \hat k)^{N/2}}\sum_{n \in \IZ^N/p \IZ^N}
\re^{{\pi \ri \hat k  n^2 \over  p}-{2\pi \ri \over p} \rho \cdot n }\prod_{1\le i<j\le N}  2 \sin \Bigl( {\pi \over \hat k  p}(\hat k (n_i-n_j) +j-i) \Bigr).
\ee
This expression for the partition function 
can be understood as a sum over saddle-points. For $U(N)$ Chern--Simons theory on $L(p,1)=\IS_3/\IZ_p$, these are 
flat connections, which can be obtained by considering embeddings of the first fundamental group into $U(N)$, modulo gauge transformations. 
Since $\pi_1\left(L(p,1)\right) = \IZ_p$ 
these embeddings are given by $N$-component vectors $n$, whose entries take values in $\IZ_p$.
However, since the residual Weyl symmetry $S_N$ of the $U(N)$ gauge group permutes the different components of $n$,  the flat connections are rather labeled by partitions $N_j$, where  $N_j$ is the number of entries in $n$ which is equal to $j-1$. The $N_j$ label the choice of vacuum
\be
U(N) \to U(N_1) \times \cdots \times U(N_p).
\ee
It is then possible \cite{akmv} to rewrite the partition function as a sum of contributions of flat connections:
\be
Z=\sum_{\{ N_j \} } Z(N_j)
\ee
where
\be
\label{fixedn2}
Z(N_j)= {(-1)^{|\Delta_+|}\over (-\ri p \hat k)^{N/2}}{1\over \prod_j N_j!} 
\sum_{w,w' \in {\cal W}} \epsilon (w) \exp
\Bigl\{ {\ri \pi \over \hat k  p} ( \rho^2 - 2\rho\cdot(\hat k 
 w'(n) +w(\rho)) + (\hat k w'(n) +w(\rho))^2
)\Bigr\}.
\ee
In this expression, the vector $n$ is any vector with $N_j$ entries equal to $j-1$. The sum over Weyl permutations $w'\in \CW$ guarantees that 
the resulting object is gauge-invariant. As shown in \cite{mcs} it is possible to rewrite (\ref{fixedn2}) as a matrix integral
\be
\label{intdef}
Z(N_k)={\re^{-{2\ri \pi \over \hat k  p} \rho^2-S_{\rm inst}(N_j)/ g_s}\over \prod_j N_j!}\int\prod_{i=1}^N {\rd x_i \over 2\pi} \, {\rm e}^{-{1\over 2 g_s} \sum_k (x_k-2\pi \ri n_k/p )^2}
 \prod_{i<j} \Bigl( 2 \sinh {x_i - x_j\over 2}
 \Bigr)^2,
\ee
where
\be
\label{cscoup}
g_s={2\pi \ri \over p \hat k}
\ee
and the pre-factor involves the gauge-theory action of each flat connection:
\be
\label{classaction}
{S_{\rm inst}(N_j)\over g_s} ={2\pi^2 \over p^2 g_s} \sum_{j=1}^p (j-1)^2 N_j. 
\ee
It can be easily checked that, although the expression (\ref{intdef}) involves a choice of vector $n$, any two choices related by a Weyl permutation lead to the 
same matrix integral. The expression (\ref{intdef}) can be regarded as a $p$-cut matrix model, where $N_k$ eigenvalues sit around the point $2\pi \ri(k-1)/p$, i.e. it 
is a matrix model with fixed filling fractions. It has in particular an asymptotic large $N$ expansion of the form
\be
\label{largen}
F_{N_k}=\log Z_{N_k} =\sum_{g=0}^\infty F_g(t_i)g_s^{2g-2}
\ee
where
\be
t_i=g_s N_i
\ee
are the partial 't Hooft parameters. In the case $p=2$, calculations around the Gaussian point of the matrix model give the genus zero free energy
\be
\label{gzerotwo}
F_0(t_1, t_2)=-{\pi^2 \over 2} t_2 + \log(-4) t_1 t_2 +F_0^{\rm G}(t_1)+F_0^{\rm G}(t_2) +F_{\rm 0}^{\rm p}(t_1, t_2).
 \ee
In this equation, the first term comes from the instanton action, the second term comes from the overall measure of the matrix integral, and $F^{\rm G}_0(t)$ 
is the Gaussian matrix model genus zero amplitude, 
\be
F_0^{\rm G}(t)={1\over 2} t^2 \Bigl( \log \, t -{3\over 4} \Bigr).
\ee
Finally, $F_{\rm 0}^{\rm p}(t_1, t_2)$ comes from fatgraphs of genus zero, and it is given by
 \be
 \ba
 F_0^{\rm p}(t_1, t_2)&={1\over 288} (t_1^4+6 t_1^3 t_2+18 t_1^2 t_2^2+6 t_1 t_2^3+t_2^4)\\
 &-{1\over 345600} (4 t_1^6+45 t_1^5 
t_2+225 t_1^4 t_2^2+1500 t_1^3 t_2^3+225 t_1^2 t_2^4+45 t_1 t_2^5+4 
t_2^6)+\cdots
\ea
\ee
For the genus one term one has:
\be
F_1(t_1, t_2) = F^{\rm G}_1(N_1) + F^{\rm G}_1(N_2)+F^{\rm p}_{1}(t_1, t_2) 
\ee
where
\be
F^{\rm G}_1(N)=\zeta'(-1)+\frac{1}{12}\log(N)
\ee
is the Gaussian matrix model contribution, and
\be
F^{\rm p}_{1}(t_1, t_2) =
-\frac{1}{288}
(t^2_1 - 6 t_1t_2 +t^2_2) + \cdots
\ee
comes from fatgraphs of genus one. Higher genus free energies can be computed analogously.

\subsection{The dual  topological string}

We will now restrict ourselves to $p=2$. The large $N$ dual model of Chern--Simons theory on $L(2,1)$ is the topological string on the anti-canonical bundle
of the Hirzebruch surface $\IF_0$.
This geometry  has  two $\IP^1$'s in $H_2(Y)$ and a compact four cycle
in $H_4(Y)$. We denote their 
associated classes  by $A_1,A_2$
and  $B$ respectively. 
The class $C = A_2 - A_1$ does not have a dual cycle in $H_4(Y)$
and  thus it  will correspond to a  non-normalizable modulus of the theory;
one can then take $A=A_2$ as  the normalizable modulus.

The mirror geometry is encoded in a family of elliptic curves $\Sigma$, which can be written as 
\be
\label{mc}
y={z_1 x^2 + x +1 -{\sqrt{(1+x + z_1 x^2)^2-4 z_2 x^2}}\over 2}. 
\ee
The periods of the one form $\lambda=\log y \rd x/x$ are given by the solutions of the associated Picard--Fuchs system:
\be
\label{pfsystem}
\ba
{\cal L}_1 &= z_2 (1-4 z_2) \xi^2_2 - 4 z_1^2 \xi_1^2 -
8 z_1 z_2 \xi_1 \xi_2 - 6 z_1\xi_1 +(1- 6 z_2) \xi_2, \\
{\cal L}_2 &= z_1 (1-4 z_1) \xi^2_1 - 4 z_2^2 \xi_2^2 -
8 z_1 z_2 \xi_1 \xi_2 - 6 z_2\xi_2 +  (1-6 z_1)\xi_1, 
\ea
\ee
where 
\be
\xi_i={\partial\over \partial z_i}.
\ee
At the large radius point $z_1=z_2=0$, the Picard--Fuchs system has two 
single logarithmic solutions, which give the mirror maps at large radius
 \be
 \label{mmap}
 \ba
 T_1&=-\log z_1 - 2 z_1 -2 z_2 - 3 z_1^2 -12 z_1 z_2 - 3 z_2^2 +\cdots,\\
  T_2&=-\log z_2 - 2 z_1 -2 z_2 - 3 z_1^2 -12 z_1 z_2 - 3 z_2^2 +\cdots.
  \ea
  \ee
  As expected, the quantity
 \be
 T_1 -T_2=\log {z_2\over z_1},
 \ee
 corresponding to the class $C$, does not receive any instanton corrections. It only depends on the ``bare parameters" of the model $z_1, z_2$ 
 and has to be regarded as a {\it parameter}. As we will see in a moment, it has a very natural meaning in the dual 
Chern--Simons theory. 

In the moduli space of this model there is another point, discovered in \cite{akmv}, which makes contact with Chern--Simons perturbation theory. This point 
is called the {\it orbifold point}, and it is defined as the point $x_1=x_2=0$ in terms of the variables:
\be \label{cov} 
  x_1=1-{z_1\over z_2}, \qquad x_2={1\over \sqrt{z_2}\left(1-{z_1\over z_2}\right)}.
  \ee
The periods near the orbifold point can be obtained again as solutions to the Picard--Fuchs system. They have the structure, 
\be
\label{speriods}
\ba
\sigma_1&=-\log(1-x_1)=\sum_{m} c_{m,0} x_1^m,\\
\sigma_2&=\sum_{m,n} c_{m,n} x_1^m x_2^n,\\
\CF_{\sigma_2}&=\sigma_2 \log(x_1)+\sum_{m,n} d_{m,n} x_1^m x_2^n \ ,
\ea
\ee
where the coefficients $c_{m,n}$ and $d_{m,n}$ are determined by the following recursions relations \cite{akmv}
\be
\label{cdrecursion}
\ba
c_{m,n}=&c_{m-1,n}{(n+2-2m)^2\over 4 (m-n) (m-1)},\cr
c_{m,n}=&{1\over n(n-1)} (c_{m,n-2} (n-m-1)(n-m-2)- c_{m-1,n-2} (n-m-1)^2),\\
d_{m,n}=&{d_{m-1,n}(n+2-2m)^2 + 4 (n+1-2m) c_{m,n}+ 4 (2m-n-2) c_{m-1,n}\over 4 (m-n) (m-1)},\cr
d_{m,n}=&{1\over n(n-1)} (d_{m,n-2} (n-m-1)(n-m-2)- d_{m-1,n-2} (n-m-1)^2 \cr &+(2n-2-2m)c_{m-1,n-2}+(2m+3-2n)c_{m,n-2}).
\ea
\ee
The orbifold point gives the large $N$ solution of Chern--Simons theory on $L(2,1)$, around a fixed, arbitray flat connection, in the sense that the perturbative topological 
string amplitudes in this frame, $F_g(\sigma_1, \sigma_2)$, correspond to the amplitudes in the $1/N$ expansion (\ref{largen}). The dictionary relating the variables 
is
\be
\ba
 t_1&={1\over 4} (\sigma_1+\sigma_2), \quad
  t_2&={1\over 4} (\sigma_1-\sigma_2).
  \ea
  \ee
Notice that the parameter of the theory
 \be
\log {z_2\over z_1} =\sigma_1=2(t_1+t_2), 
 \ee
is proportional to the {\it total} 't Hooft parameter of Chern--Simons theory $g_s N$. This is in fact required by the large $N$ duality: although $t_{1,2}$, the partial 't Hooft parameters, 
change in a natural way under the action of the 
symplectic group, the total 't Hooft parameter $t$ must be a symplectic invariant, since it is a parameter of the gauge theory. 
The derivative of the genus zero free energy of Chern--Simons theory w.r.t. the variable $s$ defined in (\ref{average}) 
is given by the following combination of the periods:
 \be
 \label{pf0}
 {\partial F_0 \over \partial s}={\pi^2\over 2} +{1\over 2}\CF_{\sigma_2}- \log(-4) \sigma_2.
  \ee
By integrating w.r.t. $\sigma_2$ and expressing the result in terms of $t_1,t_2$
one recovers the genus zero free energy of the matrix model expansion. 

In the following it will be useful to express some quantities in terms of modular forms. It turns out that \cite{abk,hkr}
\be
\tau={1\over 2\pi \ri}  {\partial^2 F_0 \over \partial s^2} 
 \ee
 is indeed a modular parameter for the curve (\ref{mc}), and in particular ${\rm Im}\, \tau >0$. The modulus
\be
\label{orbu}
u=-\frac{1}{8\sqrt{1-x_1}}x^2_1x_2^2+\frac{1}{2}\left( \sqrt{1-x_1}+\frac{1}{\sqrt{1-x_1}} \right),
\ee
is closely related to the modulus of a Seiberg--Witten curve \cite{sw} with
\be
\tau_{\rm SW}=2\tau+1,
\ee
and it can be expressed in terms 
of elliptic theta functions:
\be
\label{eorbu}
u=\frac{\vartheta^4_4-\vartheta^4_2}{\vartheta^4_3}(2\tau)=1-32 q+256 q^2+\cdots
\ee
where $q=\re^{2 \pi i \tau}$. Notice that the orbifold point in the moduli space of local $\IF_0$ corresponds to $u=1$, i.e. to the monopole point of the Seiberg--Witten curve. 

Finally one can check that the genus one free energy is given, in terms of modular forms, by 
\be
\label{f1}
F_1=-\log\, \eta(\tau_{\rm SW})+2 \zeta'(-1)+\frac{1}{6}\log{g_s\over 4},
\ee
In particular, we find
\be
F^{\rm p}_{1}(t_1, t_2) = - \log \eta(\tau_{\rm SW} ) +
\frac{1}{12}\log\left(
\frac{t_1t_2}{16}\right).
\ee

The partition function of Chern--Simons theory on $L(2,1)$ can be written in terms of a two-cut matrix model \cite{mcs,akmv}. The cuts are centered around $x=0$, $x=\pi \ri$, and they are 
precisely the cuts which appear in the mirror curve (\ref{mc}). If we write them as
\be
(-a, a), \qquad (\pi \ri -b, \pi \ri +b), 
\ee
in terms of the endpoints $a, b$, it is easy to see that they are determined by the equations (see also \cite{hy})
\be
\ba
\cosh \, a&={1\over 2 {\sqrt{ z_1}}}+{\sqrt{z_2\over z_1}}=1 + 2 t_1 +  t_1 (t_1+t_2) + \cdots,\\
\cosh \, b&=-{1\over 2 {\sqrt{ z_1}}}+{\sqrt{z_2\over z_1}}=1 +2 t_2 +  t_2 (t_1+t_2) + \cdots .
\ea
\ee

\subsection{Two special slices}

The above results simplify considerably along particular submanifolds in moduli space, which will be relevant in the analysis of the phase 
structure in the next section. 

The {\it slice $t_2=0$} corresponds, from the point of view of the 
gauge theory, to an expansion around the trivial flat connection. Up to a trivial rescaling of the coupling constant $g_s$,  the theory 
reduces to Chern--Simons theory on the three-sphere $\IS^3$. From the 
point of view of topological string theory, the theory is equivalent to the resolved conifold. The genus zero 
and genus one free energies are given by, 
\be
\ba
F^{\IS^3}_0(t)&={t^3\over 12} -{\pi^2 t \over 6}-{\rm Li}_3(\re^{-t})+\zeta(3),\\
F^{\IS^3}_1(t)&=-{t\over 24}-{1\over 12} \log (1-\re^{-t}) +{1\over 12} \log t + \zeta'(-1) -{1\over 12} \log N.
\ea
\ee
The instanton action (\ref{hata}) on $t_1=t$, $t_2=0$ can be evaluated in closed form (see the Appendix) and it is given by
\be
\label{ainst}
 A(t)= 2 \, {\rm Li}_2(\re^{-t/2}) -2\,  {\rm Li}_2(-\re^{-t/2}).
\ee

The other relevant slice, for us, corresponds to the {\it symmetric slice} $t_1=t_2$. In terms of the orbifold coordinates, this slice is parametrized by
\be
x_2=\sigma_2=0, \quad x_1=1-\re^{-\sigma_1}.
\ee
On this slice we have the following exact expressions for the prepotential 
and for the modular parameter in terms of the  flat coordinate $\sigma_1$:
\be
\label{f0}
F_0(\sigma_1)=-{1\over 8} {\rm Li}_3(\re^{-\sigma_1})+ {\sigma_1^3 \over 96}+{\pi \ri  \over 16}\sigma_1^2-{7\pi^2 \over 48} \sigma_1+{\zeta(3)\over 8},
\ee
\be
\label{tau}
\tau= \ri \frac{ K'(k)}{ K(k)} -1, \qquad k^2=1-\re^{-\sigma_1},
\ee
where $K(k)$ is the standard elliptic integral. The proof of the above identities is given in the Appendix.

\sectiono{Large $N$ phase diagram}

So far we have just recalled the {\it perturbative} large $N$ duality 
between Chern--Simons theory on $L(2,1)$ expanded around a particular flat connection (corresponding to a choice of filling fractions $N_1,N_2$), and topological string theory 
on local $\IF_0$ with K\"ahler parameters $t_1,t_2$.

However, as we explained in the introduction, when studying the large $N$ expansion of the non-perturbative partition function, the choice of filling fractions  
(dominant saddle) is not arbitrary but it is dictated by the value of the total 't Hooft coupling $t$. We will now study the large $N$ phase diagram for the Chern--Simons theory on $L(2,1)$, i.e. 
we will determine the saddle which gives the dominant contribution to the non-perturbative partition function as 
we move on the complex plane of $t=g_s N$ (equivalently, on the complex plane of $g_s$).

We first notice two exact symmetries of the non-perturbative partition function. If we conjugate $g_s$, we find
\be
\label{conjugate}
Z(N,g_s^*)=(-1)^{c(N)} Z^*(N,g_s), 
\ee
where $c(N)$ is a half-integer that only depends on $N$. This symmetry is valid for all $p$. The second symmetry is related to changing $g_s \rightarrow -g_s$. If we write 
\be
\label{sumn2}
Z(N, g_s) =\sum_{N_2=0}^N Z(N,N_2, g_s)
\ee
we have that
\be
\label{signchange}
Z(N,N_2,-g_s)=(-1)^{c'(N)} {\rm e}^{-\frac{\pi^2 N^2}{2t}} (-1)^{(N-1)N_2} Z(N,N-N_2,g_s),
\ee
where $c'(N)$ is again a half-integer that only depends on $N$. 

In determining the phase structure, we are interested in knowing which is the $N_2$ which contributes the most to the non-perturbative partition function (in absolute value). As we will see, 
for $N$ large, 
the sum in (\ref{sumn2}) will be very peaked around a particular saddle value $(N_1^*, N_2^*)$. We see, from (\ref{conjugate}), that conjugating $g_s$ does not change this saddle, 
while from (\ref{signchange}) we see that the phase diagram is invariant under the simultaneous change
\be
g_s \leftrightarrow -g_s, \qquad (N_1^*, N_2^*) \leftrightarrow (N_2^*, N_1^*).
\ee
In practice, this means that in order to determine the phase diagram, we can restrict ourselves to the first quadrant of the complex $g_s$ plane
\be
0\le {\rm Arg}(g_s) \le {\pi \over 2}. 
\ee
In order to proceed, we first study the limiting cases of real, positive $g_s$ and purely imaginary $g_s$.

\subsection{Real $g_s$: the boundary expansion}

The classical action of the flat connection (\ref{classaction}) has the value 
\be
S_{\rm inst}(N_2) ={\pi^2 \over 2} N_2.
\ee
When $g_s$ is real and positive, it gives an exponential suppression for all the non--trivial flat connections $N_2 >0$. Therefore, at least 
when the 't Hooft coupling is sufficiently small, the dominant saddle point  is expected  to be at the boundary 
$(N_1^*, N_2^*)=(N,0)$. In fact, it is easy to see that for any real, positive $t$, 
this boundary point is the dominant saddle. To prove it, it suffices to show that the condition (\ref{stableboundary}) 
holds on this slice, where $A(t)$ is the large $N$ instanton action corresponding to the first non-trivial instanton sector with $N_2=1$. This action is given in (\ref{ainst}), 
and it resums quantum fluctuations 
around the classical instanton solution. At small $t$ it should have the general structure \cite{gm} 
\be
\exp\left( -{A(t) \over g_s} \right) \sim \biggl( {c_2 \over g_s}\biggr)^{ c_3 N}{\rm e}^{-{c_1/ g_s}}\ee
where the prefactor is the one-loop fluctuation around the classical instanton with action $c_1=\pi^2/2$, $c_3 N$ is the number of zero modes at large $N$, and 
$c_2$ is a numerical constant. Therefore, the expansion of $A(t)$ around $t=0$ should be of the form 
\be
A(t) = c_1 - c_3 t \log \left( {c_2 \over t} \right)+ \CO(t).
\ee
Indeed, we find from the exact expression (\ref{ainst}) that
\be
A(t)={\pi^2 \over 2} -t \log \left( {4\over t}\right) -t +\CO(t^3). 
\ee
This large $N$ instanton action is real for $t\in [0, \infty)$ and it decreases monotonically from 
\be
A(0)={\pi^2 \over 2} \quad {\text {to}} \quad \lim_{t\to \infty} A(t)=0.
\ee
In particular it does not vanish for any finite $t$. This can be seen in \figref{numinst}, where the continuous line represents $A(t)/t$ as a function of $t/\pi$. 
As we explain in the Appendix, this is equivalent to the absence of 
phase transition for $q$-deformed 2d Yang--Mills with $p=2$ \cite{sara,abms,jm}. We conclude that, for real $g_s$, the large $N$ asymptotics of the partition function of 
Chern--Simons theory on $L(2,1)$ is simply given by the $1/N$ expansion of Chern--Simons theory on $\IS^3$, 
\be
\label{realas}
\log Z_{L(2,1)}(N,g_s) =\sum_{g=0}^{\infty} g_s^{2g-2} F_g^{\IS^3}(t).
\ee
In other words, the relevant saddle geometry along this direction in moduli space is just the resolved conifold geometry. 

Of course, the result (\ref{realas}) 
has an infinite number of exponentially suppressed instanton corrections of the form (\ref{npboundary}). From the point of view of {\it classical} asymptotic analysis, these subleading saddles 
are not taken into account, but of course they are important thanks, among other things, to the Stokes phenomenon that we will now uncover. 
 
\begin{figure}[!ht]
\leavevmode
\begin{center}
\includegraphics[height=4cm]{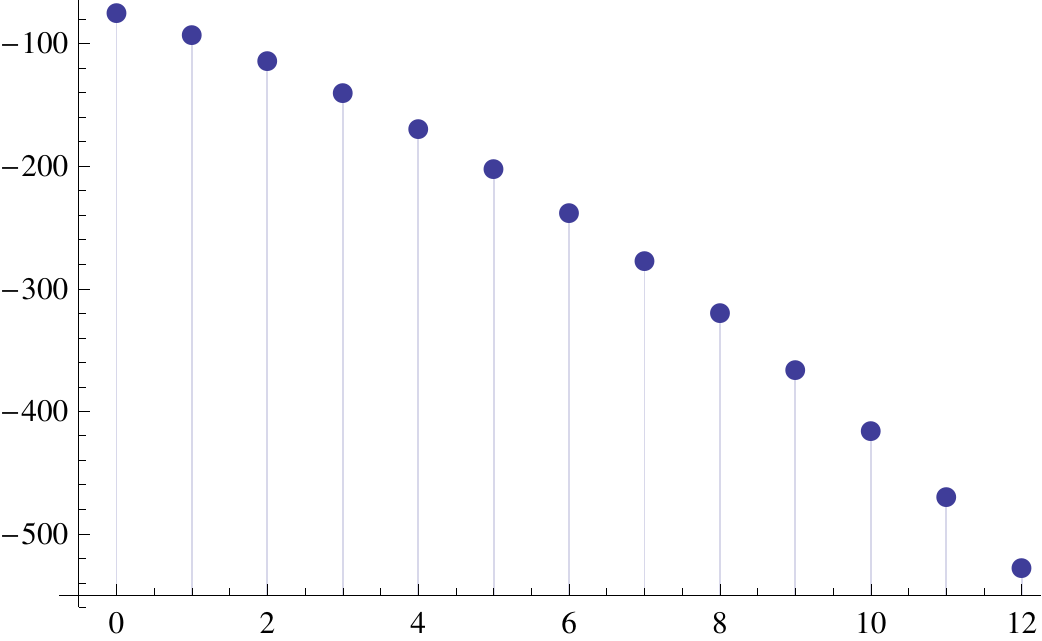}\qquad 
\includegraphics[height=4cm]{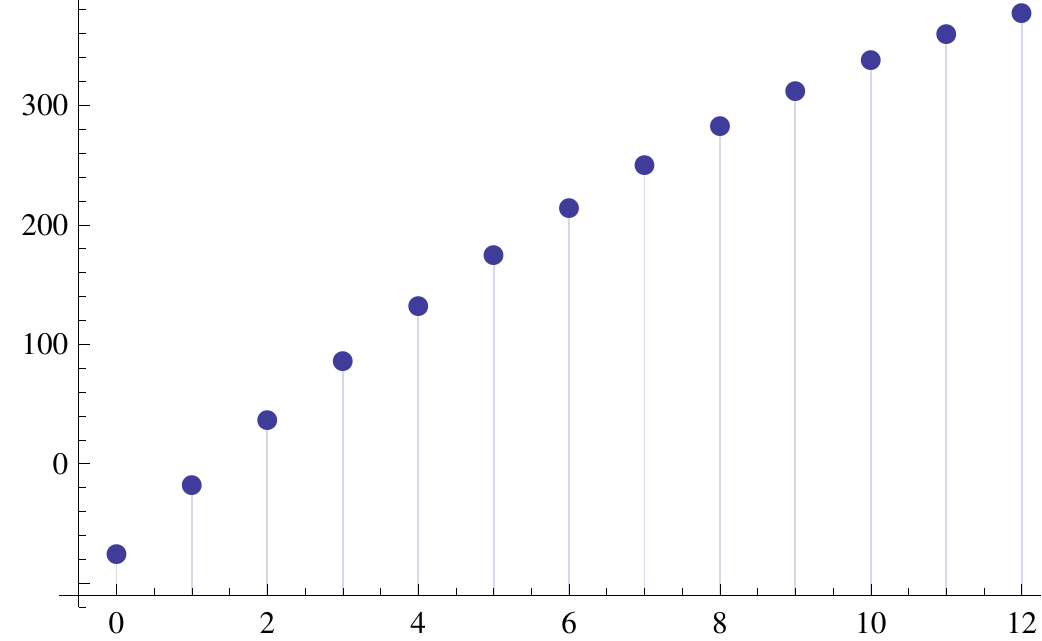}
\end{center}
\caption{In this figure we plot $\log |Z(N, N_2, g_s)|$ as a function of $N_2=0, \cdots, N$, for $N=12$ and for $g_s=\pm\pi/24$ 
respectively on the left and on the right.
The dominant configurations are 
 $(N_1^*,N_2^*)=(N,0)$ and $(0,N)$, respectively,
  in agreement with the symmetry (\ref{signchange}).
}
\label{peack1}
\end{figure} 

The above analysis can be confirmed numerically for small values of $N$. In general, a simple way to estimate the leading saddle $(N_1^*, N_2^*)$ for finite $N$ is to 
calculate the partition functions $Z(N, N_2, g_s)$ for fixed values of $N_2$, and see which one is the largest in absolute value, and therefore gives 
the dominant contribution in (\ref{sumn2}). In \figref{peack1} 
we plot $\log |Z(N, N_2, g_s)|$, for $g_s$ real and positive. As expected, the 
dominant configuration is $(N_1^*,N_2^*)=(N,0)$ ($(N_1^*,N_2^*)=(0,N)$  for $g_s$ real and negative).

\begin{figure}[!ht]
\leavevmode
\begin{center}
\includegraphics[height=8cm]{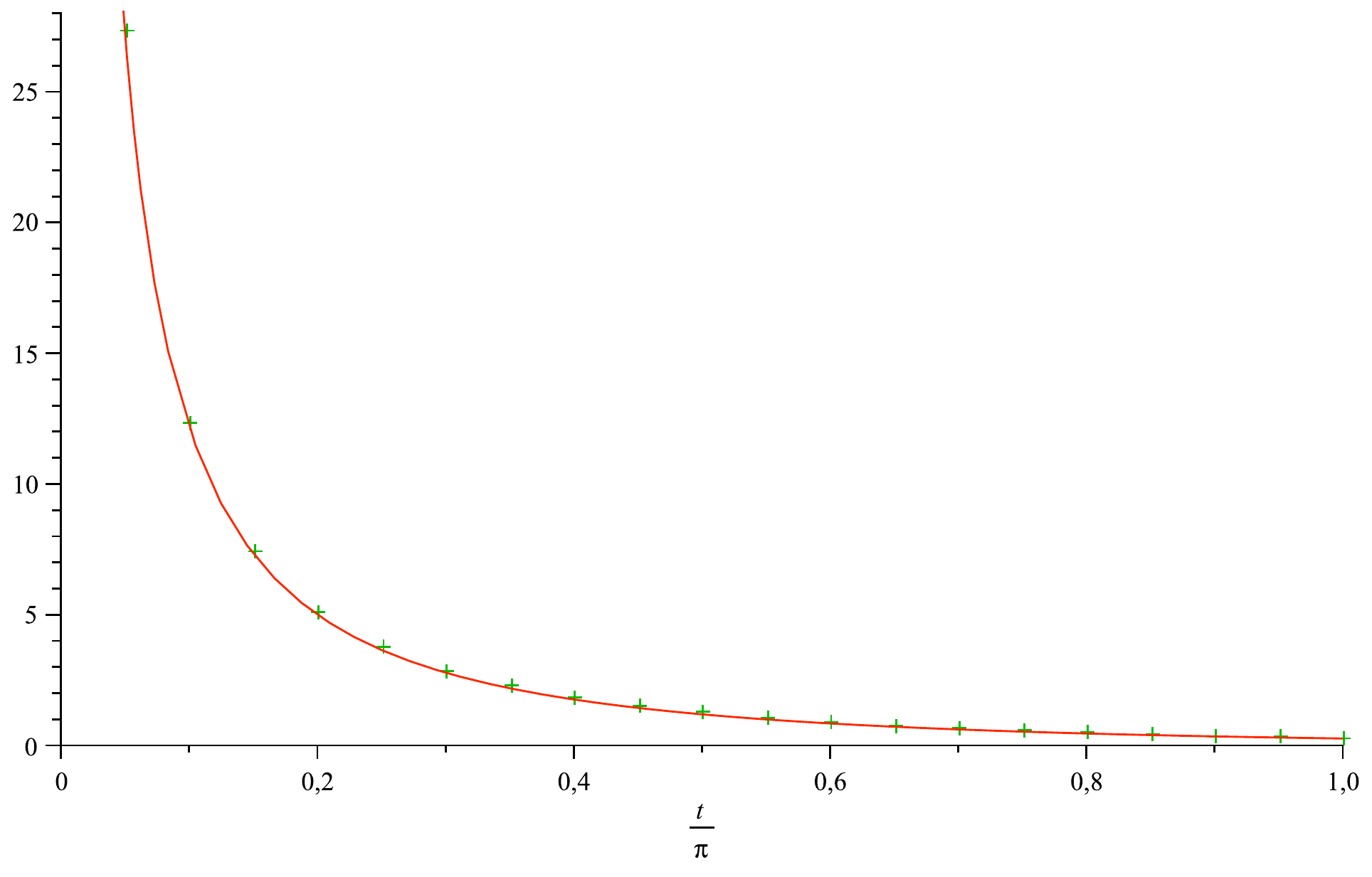}
\end{center}
\caption{In this figure, the crosses represent the right hand side of (\ref{iasn}), for $N=24$ and different values of $t$, while the continuous line represents the 
value of $A(t)/t$ obtained from (\ref{ainst}).}
\label{numinst}
\end{figure} 

We can also confirm numerically the value of the instanton action (\ref{ainst}). From (\ref{npboundary}) we find the large $N$ asymptotics
\be
\label{iasn}
-{1\over N} \log \left( {Z(N, N_2=1, g_s) \over Z(N, N_2=0, g_s)}\right)= {A(t) \over t} + \CO\left( {1\over N } \right). 
\ee
 In \figref{numinst} we plot the values of the r.h.s. for $N=24$ and different values of $t=N g_s$, and we compare them to the value of $A(t)/t$ given by (\ref{ainst}). As we can see, the agreement is excellent, and it can be further improved by extracting the subdominant tails in (\ref{iasn}) through the technique of Richardson transforms.

\subsection{Imaginary $g_s$: the interior expansion}

\begin{figure}[!ht]
\leavevmode
\begin{center}
\includegraphics[height=4cm]{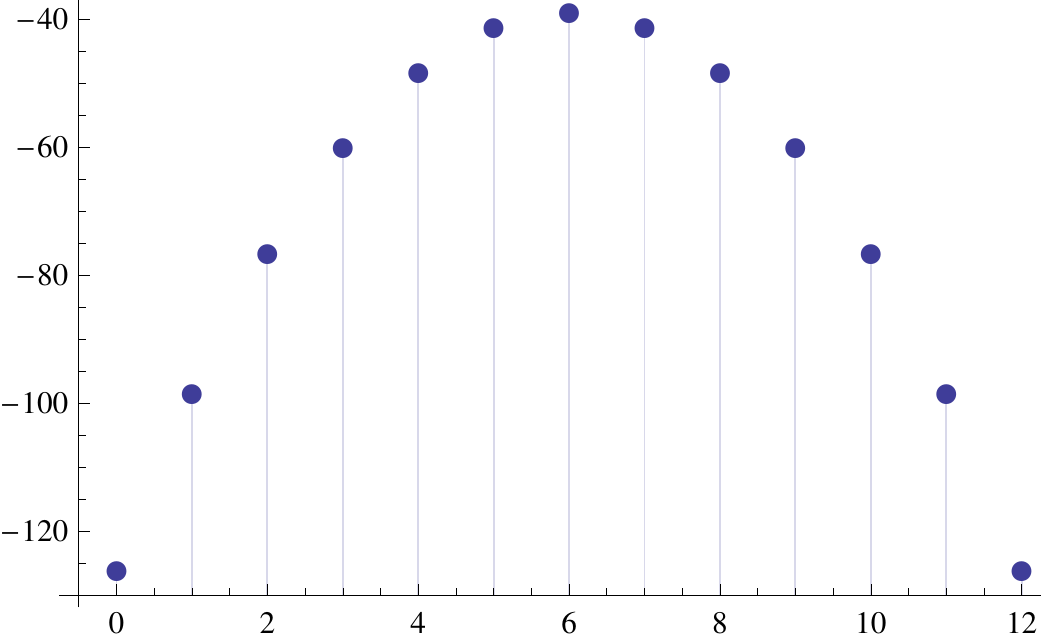}\qquad \includegraphics[height=4cm]{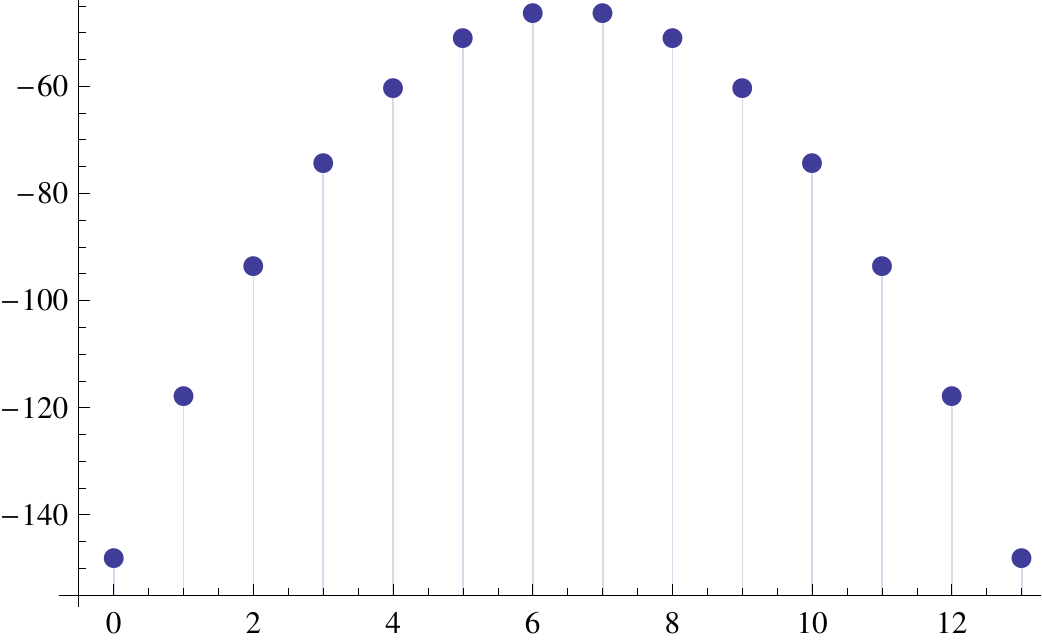}
\end{center}
\caption{In this figure we plot $\log |Z(N, N_2, g_s)|$ as a function of $N_2=0, \cdots, N$, on the left  for $N=12$ and $g_s=\pi \ri /48$,  on the right  for $N=13$
 and $g_s=\pi \ri/52$.
 The largest value is obtained when $N_2={N/2}$ for $N$ even and when $N=(N\pm 1)/2$ for $N$ odd.}
\label{peack2}
\end{figure}

Let us analyze now the large $N$ asymptotics along the direction ${\rm Arg}(g_s)=\pi/2$. The boundary point $(N,0)$ can not be a solution any longer, since (\ref{stableboundary}) 
is no longer verified (at least for small $t$). This is easy to understand: for imaginary $g_s$, (\ref{classaction}) is purely imaginary and the non-trivial instanton sectors are 
not suppressed anymore. We then have to look for saddles satisfying the condition (\ref{realvanish}), which in this case reads:
\be
{\rm Re}\Big[{1\over g_s}\Bigl({\partial F_0 \over \partial t_1} -{\partial F_0 \over \partial t_2} \Bigr)\Big]=0.
\ee
It is easy to see that the real part of $F_0(t_1, t_2)$ is symmetric under the exchange of $t_1$ and $t_2$, therefore the configuration 
\be
\label{intsaddle}
t_1=t_2={t\over 2},
\ee
which is an interior point, is a saddle. We claim that this gives in fact the dominant saddle when $g_s$ is imaginary. Of course, there could be other saddles which actually dominate 
the asymptotics. A first indication that this is not the case, and that (\ref{intsaddle}) is the relevant saddle, 
comes from the numerical analysis of $Z(N, N_2, g_s)$ for small $N$. As we see in \figref{peack2}, for $N$ even 
the largest contribution comes from $N_2=N/2$, while for $N$ odd the dominant contributions have $N_2=(N\pm 1)/2$. 
 
 According to the general discussion in section 2, since the saddle (\ref{intsaddle}) is an interior point, 
the large $N$ asymptotics of the partition function should be given by (\ref{npinterior}). In our case, we have
\be
\epsilon={1\over 2},\qquad \mu=\nu=0, \qquad {u \over g_s}={\pi^2\over 2g_s}=-{\pi \ri \hat k \over 2}. 
\ee
The resulting theta function depends on the value of $N$ modulo two. If $N$ is {\it even}, $N\epsilon$ is an integer which can be reabsorbed in a shift of $n\in \IZ$, which is summed over. Therefore, (\ref{biget}) is given by
\be
\label{et}
\vartheta_e=\sum_{n\in \IZ} \exp\left( \pi \ri \tau n^2+\ri {\hat k \pi \over 2} n\right).
\ee
Notice that, since $g_s$ is imaginary, $k$ is real, and the second term leads to an oscillatory behavior. If $N$ is {\it odd}, $N\epsilon$ is a half-integer. We can absorb its integer part 
by a shift of $n$  and we obtain for (\ref{biget})
\be
\label{ot}
\vartheta_o=\sum_{n\in \IZ+{1\over 2} } \exp \left(\pi \ri \tau n^2+\ri {\hat k \pi \over 2} n\right).
\ee

We then claim that, when $g_s$ is  imaginary, the asymptotic behavior of $F(N,g_s)=\log Z(N,g_s)$, the free energy of Chern--Simons theory on $L(2,1)$, at large $N$ and fixed 't Hooft 
coupling, is given by the logarithm of (\ref{npinterior}). At leading order we have
\be
F(N,g_s) = g_s^2 F_0\left({t\over 2}, {t\over 2}\right) +F_1\left({t\over 2}, {t\over 2}\right)+ \log \Theta_{0,0}(u,\tau)+\cdots
\ee
where $t=Ng_s$. Notice that
\begin{enumerate}
\item The perturbative free energies $F_g(t/2, t/2)$ are the genus $g$ free energies of topological string theory on $\IF_0$ on the slice $t_1=t_2=t/2$. For genus zero, it is given by the 
explicit expression (\ref{f0}). The higher genus amplitudes are essentially quasi-modular forms of the modular parameter $\tau$ \cite{abk,hkr} which is given in (\ref{tau}) by an explicit function of 
\be
\sigma_1={2\pi \ri N \over k+N},
\ee
where we used the original Chern--Simons parameters. In particular, $F_1$ is given in (\ref{f1}).

\item The asymptotics depends on the parity of $N$ through the theta function, which is given by (\ref{et}), (\ref{ot}) for $N$ even and odd, respectively. 

\end{enumerate}

In the quantum Chern--Simons gauge theory, $\hat k=k+N$ is an integer, and $g_s$ is purely imaginary. Therefore, the above asymptotics is in principle 
the relevant one to understand the large $N$ 't Hooft limit of the quantum Chern--Simons invariant of $L(2,1)$. 

\begin{figure}[!ht]
\leavevmode
\begin{center}
\includegraphics[height=4cm]{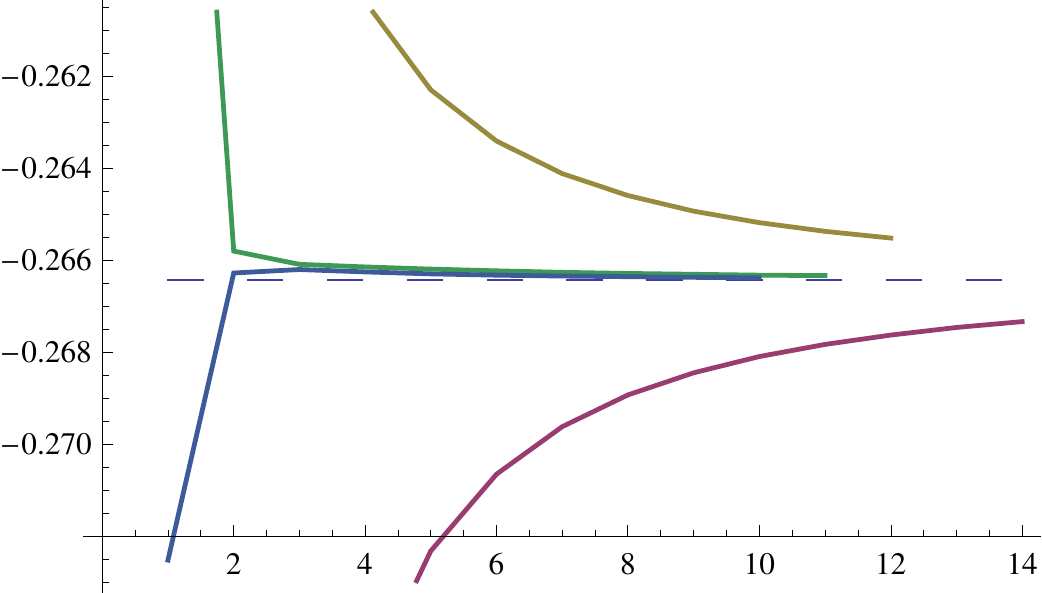}\quad 
\includegraphics[height=4cm]{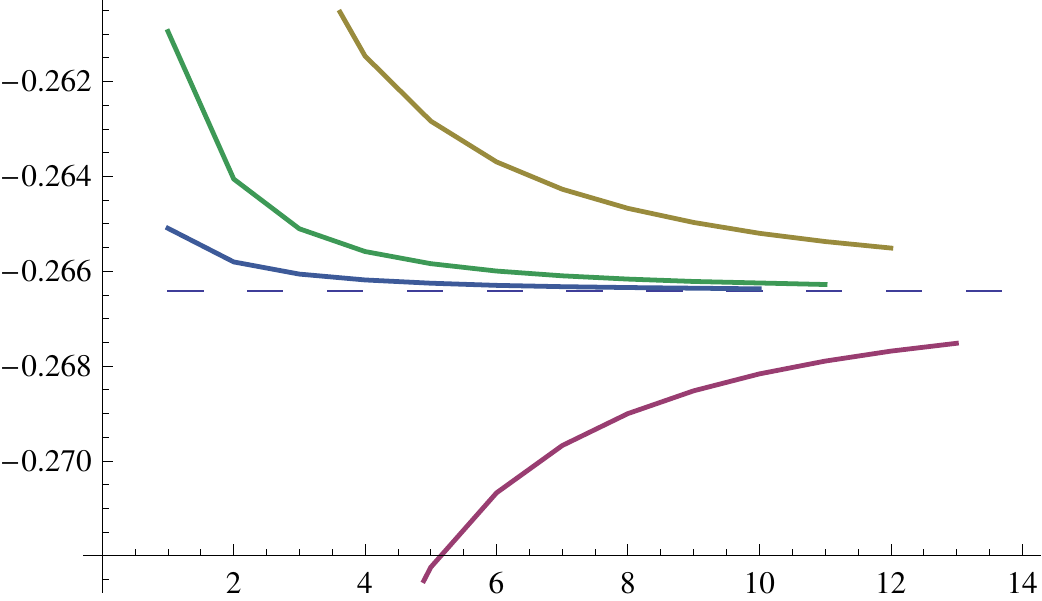}
\end{center}
\begin{center}
\includegraphics[height=4cm]{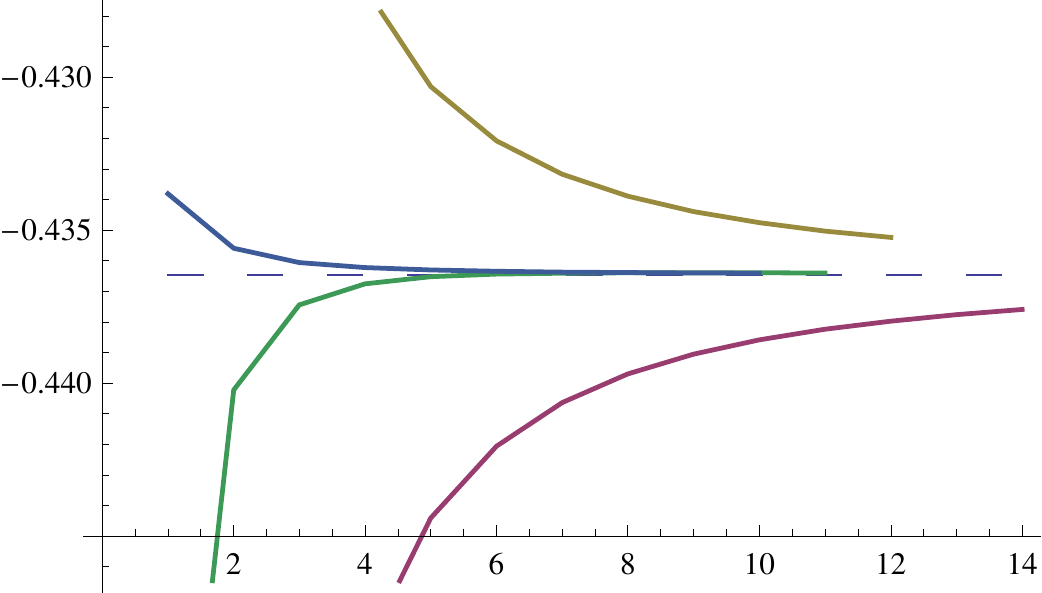}\quad 
\includegraphics[height=4cm]{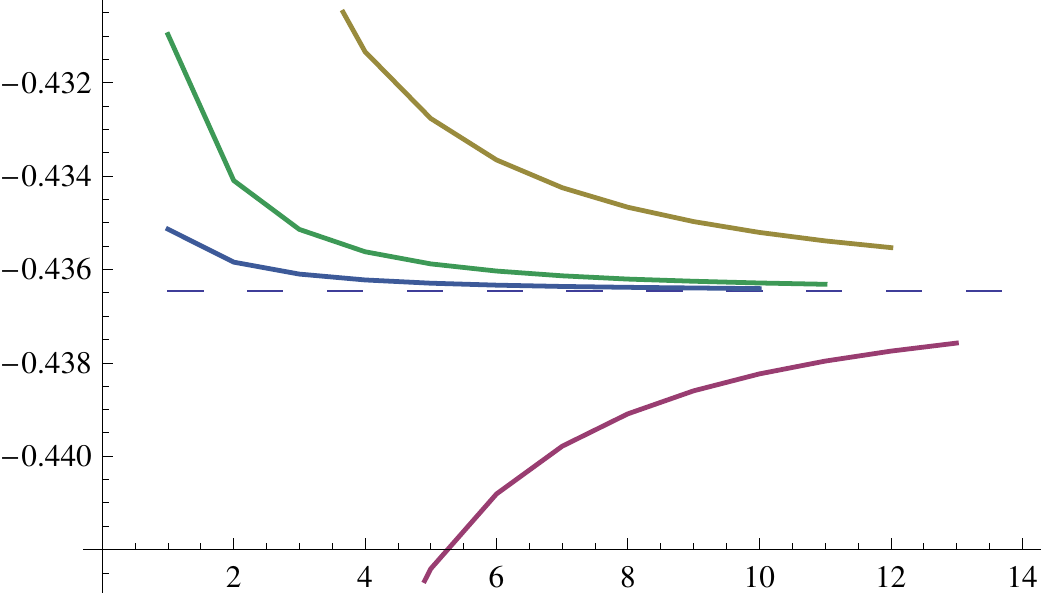}
\end{center}
\caption{On the left column we plot the sequence ${\rm Re}\, f_{e}(\ell,z)$
as well as its Richardson transforms ${\rm Re}\, f^{(k)}_{e}(\ell,z)$
for $k=1,2,3$, and for $\ell=2\cdots 14$.
On the right column we  plot the sequence ${\rm Re}\, f_{o}(\ell,z)$
as well as its Richardson transforms ${\rm Re}\, f^{(k)}_{o}(\ell,z)$
for $k=1,2,3$ for $\ell=2\cdots 14$. The plots on the top are for $z=1/4$, and the 
plots of the bottom are for $z=1/8$. 
The dashed lines in the plots
show  the expected value $-{\rm Re}\, F_0( z)/(\pi z)^2 $.}
\label{exf0}
\end{figure} 

Our claim about the asymptotics follows from the identification of (\ref{intsaddle}) as the relevant saddle point, and from the general theory reviewed in section 2 concerning instanton 
corrections. We now provide detailed numerical evidence for our claim. Instead of working with $N$ and $g_s$ we will use $N$ and $z$, which is defined by
\be 
z=\frac{N}{\hat k}.
\ee
Let us now define the sequences of complex numbers obtained from the non-perturbative partition function with $N=2\ell,2\ell+1$ respectively:
\be
f_e(\ell,z)={1\over (2\ell)^2}  \log Z (2\ell,z), \qquad f_o(\ell,z)={1\over (2\ell+1)^2}\log Z (2\ell+1,z).
\ee 
According to our claim, these sequences have the large $\ell$ asymptotics 
\be
\ba
f_e(\ell,z)&=-\frac{F_0(z)}{(\pi z)^2 }+{1\over (2\ell)^2} \left( F_1(z)+\log\vartheta_e(z)\right) +\mathcal{O}(\ell^{-4}),\\ 
f_o(\ell,z)&=-\frac{F_0(z)}{(\pi z)^2 }+{1\over (2\ell+1)^2} \left( F_1(z)+\log\vartheta_o(z)\right) +\mathcal{O}(\ell^{-4})
\ea
\ee
where $F_{0,1}(z)$ are the genus $0,1$ free energies evaluated on the slice (\ref{intsaddle}). Notice that the theta functions have in general oscillatory behavior as a function of 
$\ell$, but this can be fixed by choosing special values of $\hat k$. For example, if $\hat k=0$ mod $4$ we have:
\be
\vartheta_e(z)=\vartheta_3(\tau), \qquad  
\vartheta_o(z)=\pm \vartheta_2(\tau),
\ee
where the sign depends on the value of $\hat k$ mod $8$. In this way, the resulting large $\ell$ asymptotics is simply a series in inverse powers of $\ell$, and we can use standard techniques of accelerated convergence to compare the actual values of $f_{e,o}(\ell,z)$, computed from the exact partition function of Chern--Simons theory for low $\ell$, to the predicted asymptotics.

\begin{figure}[!ht]
\leavevmode
\begin{center}
\includegraphics[height=4cm]{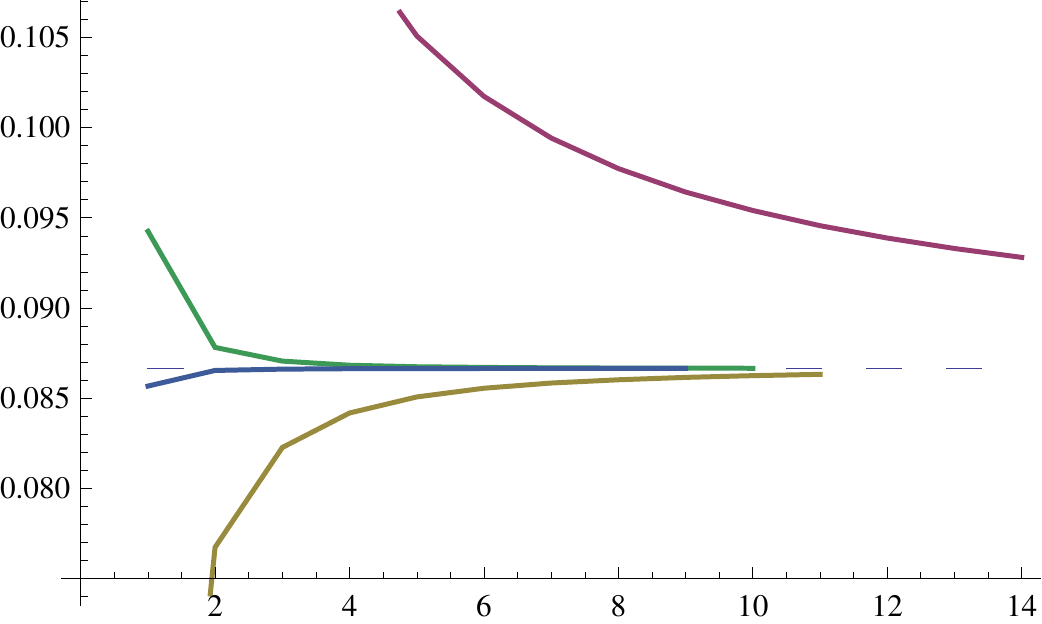}\quad \includegraphics[height=4cm]{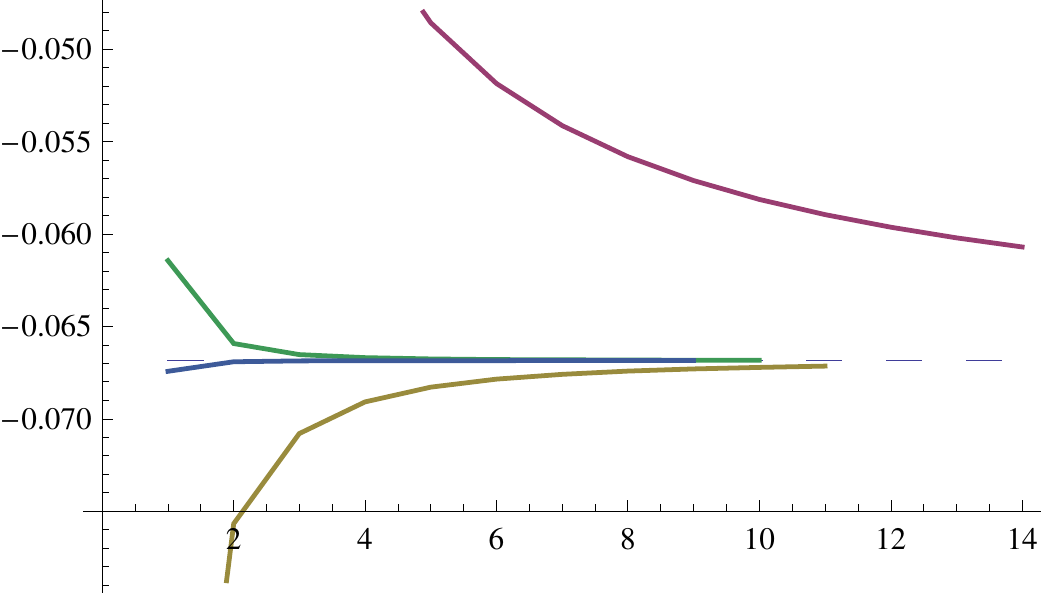}
\end{center}
\caption{The sequence ${\rm Re}\, \Theta_{eo}(z)$ for $z=1/4,1/8$, together with its Richardson transforms ${\rm Re}\, \Theta^{(k)}_{eo}(z)$ for $k=1,2,3$.
The dashed  lines in both plots
show  the expected value ${\rm Re}(\log\vartheta_e(z)-\log\vartheta_o(z))$.
}
\label{extheta}
\end{figure}

In practice, the computation of (\ref{znp}) involves generating a large number of configurations for the vectors $n$, and we obtained their numerical values for $N=1, \cdots, 30$ and for different 
values of $z$. This means that we obtained the values of the sequences $f_{e,o}(\ell,z)$ up to $\ell=14$. 
Once these sequences are computed, we obtain their Richardson transforms (see 
for example \cite{bo})  
\be
f_{e,o}^{(r)}(\ell,z)=\sum_{k\geq 0}{f_{e,o}(\ell+k,z)(\ell+k)^r(-1)^{k+r}\over k!(r-k)!} 
\ee
in order to accelerate the convergence. In \figref{exf0} we plot the sequences ${\rm Re}\, f_{e,o}(\ell,z)$
as well as their  Richardson transforms ${\rm Re}\, f^{(k)}_{e,o}(\ell,z)$
for $k=1,2,3$ for different values of $z$ together with their 
 expected value $-{\rm Re}\, F_0( z)/(\pi z)^2 $.
As we can see, the agreement is excellent. 

\begin{figure}[!ht]
\leavevmode
\begin{center}
\includegraphics[height=4cm]{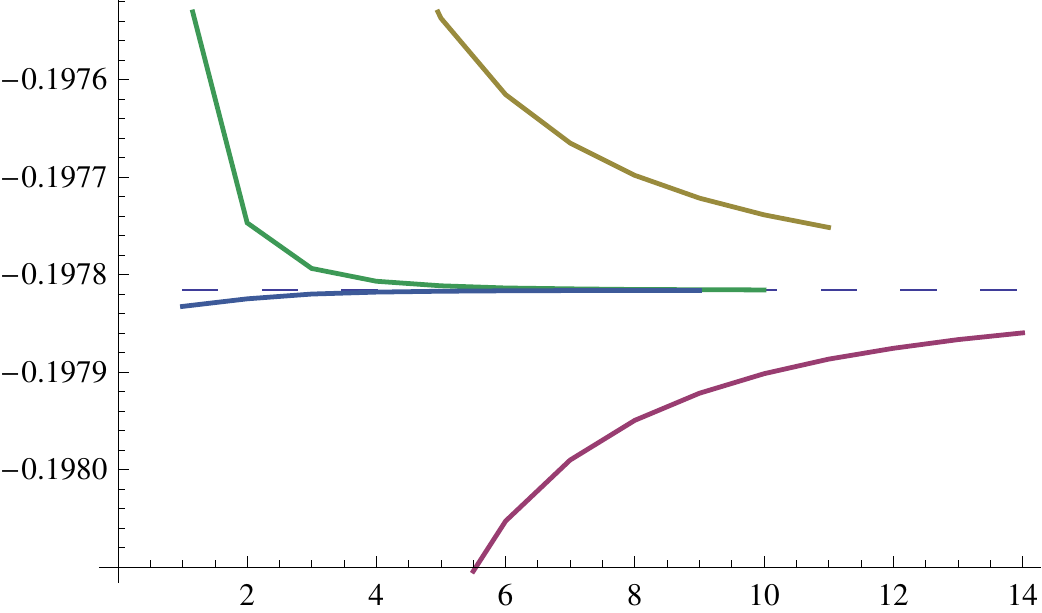} \quad 
\includegraphics[height=4cm]{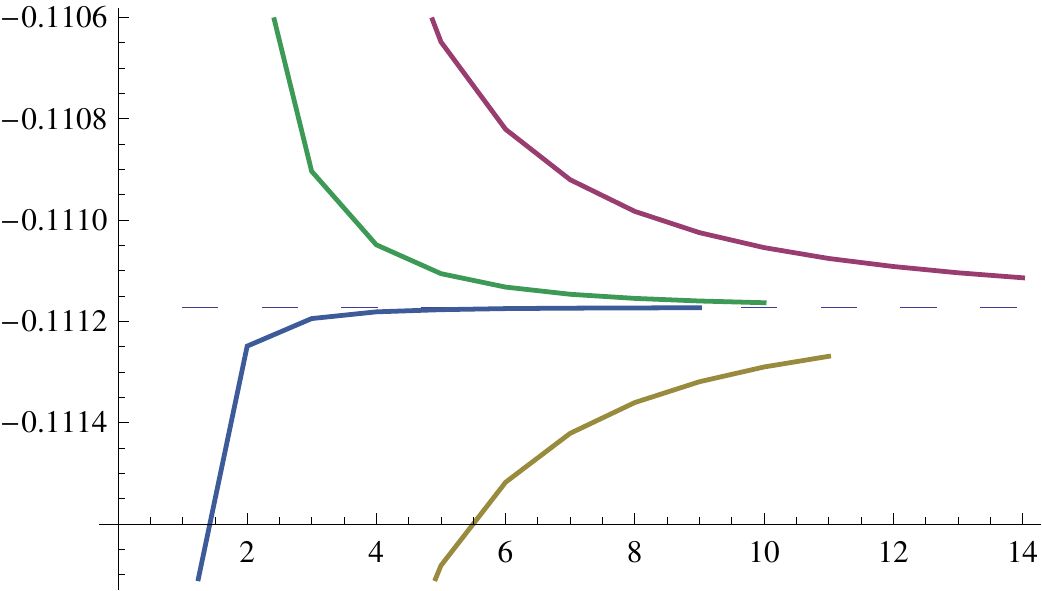}\quad 
\end{center} \begin{center}
\includegraphics[height=4cm]{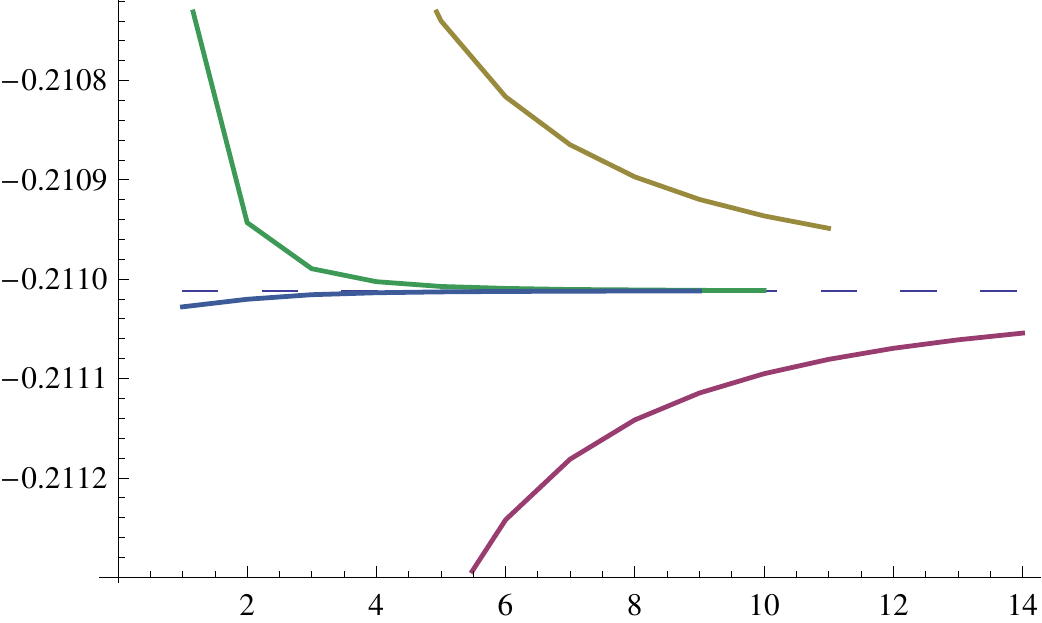}\quad
\includegraphics[height=4cm]{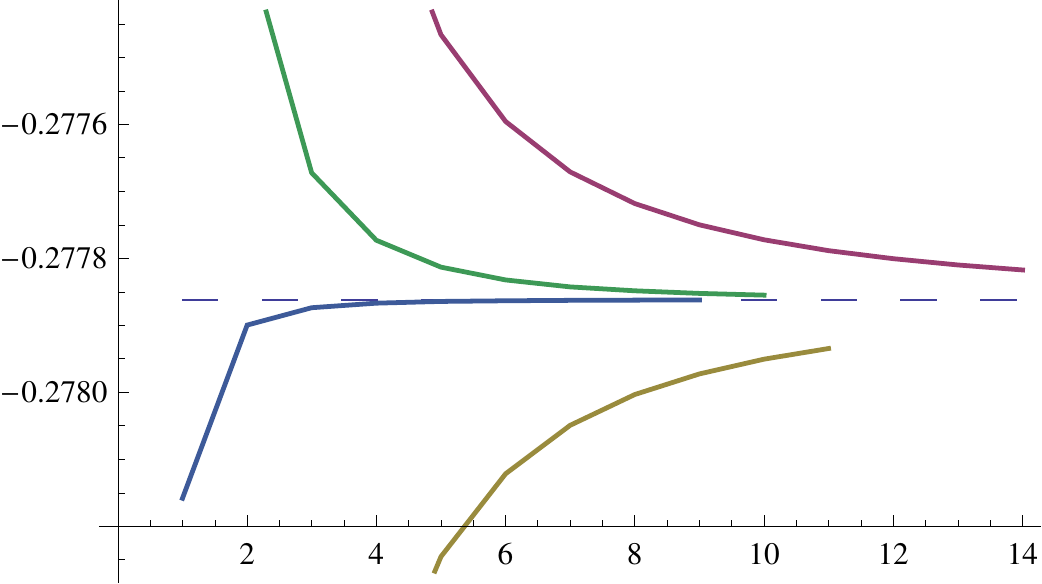}
\end{center}
\caption{
On the left column we plot the sequence ${\rm Re}\, f_{e,1}(\ell,z)$
as well as its Richardson transforms ${\rm Re}\,f^{(k)}_{e,1}(\ell,z)$
for $k=1,2,3$, and for $\ell=2\cdots 14$.
On the right column we plot the sequence ${\rm Re}\,f_{o,1}(\ell,z)$
as well as its Richardson transforms ${\rm Re}\,f^{(k)}_{o,1}(\ell,z)$, again 
for $k=1,2,3$ and for $\ell=2\cdots 14$. The plots on the top are for $z=1/4$, and the 
plots of the bottom are for $z=1/8$. 
The dashed line
shows  the expected value ${\rm Re}\,(F_1(z)+\log\vartheta_{e,o}(z))$.
}
\label{subleading}
\end{figure} 

In order to test the subleading behavior, we define the following sequences:
\be
\ba
\Theta_{eo}(\ell,z)&= (2\ell)^2 (f_e(\ell,z)- f_o(\ell,z)),\\
f_{e,1}(\ell,z)&=(2\ell)^2\left(f_e(\ell,z)+\frac{F_0(z)}{(\pi z)^2 }\right), \\
f_{o,1}(\ell,z)&=(2\ell+1)^2 \left( f_o(\ell,z) +\frac{F_0(z)}{(\pi z)^2 }\right),
\ea
\ee
with the expected asymptotic behavior:
\be
\ba
\lim_{\ell \to\infty}\Theta_{eo}(\ell,z)&=\log\vartheta_e(z)-\log\vartheta_o(z)\\
\lim_{\ell \to\infty}f_{e,1}(\ell,z)&= F_1(z)+\log\vartheta_e(z),  \\
\lim_{\ell \to\infty}f_{o,1}(\ell,z)&=F_1(z)+\log\vartheta_o(z).
\ea
\ee
As we can see in \figref{subleading}, the agreement between the analytic prediction and the actual asymptotic behavior is again extremely good. This confirms 
our identification of the saddle point (\ref{intsaddle}), as well as the formalism of \cite{bde,eynard,em} to incorporate instanton 
corrections in the oscillatory case. Although these corrections are subleading in the case of the free energy, they can appear at leading order 
when studying expectation values of operators like Wilson loops, as noticed in \cite{bde} in the context of matrix models.

\subsection{The phase diagram for complex $g_s$}

\begin{figure}[!ht]
\leavevmode
\begin{center}
\includegraphics[height=7cm]{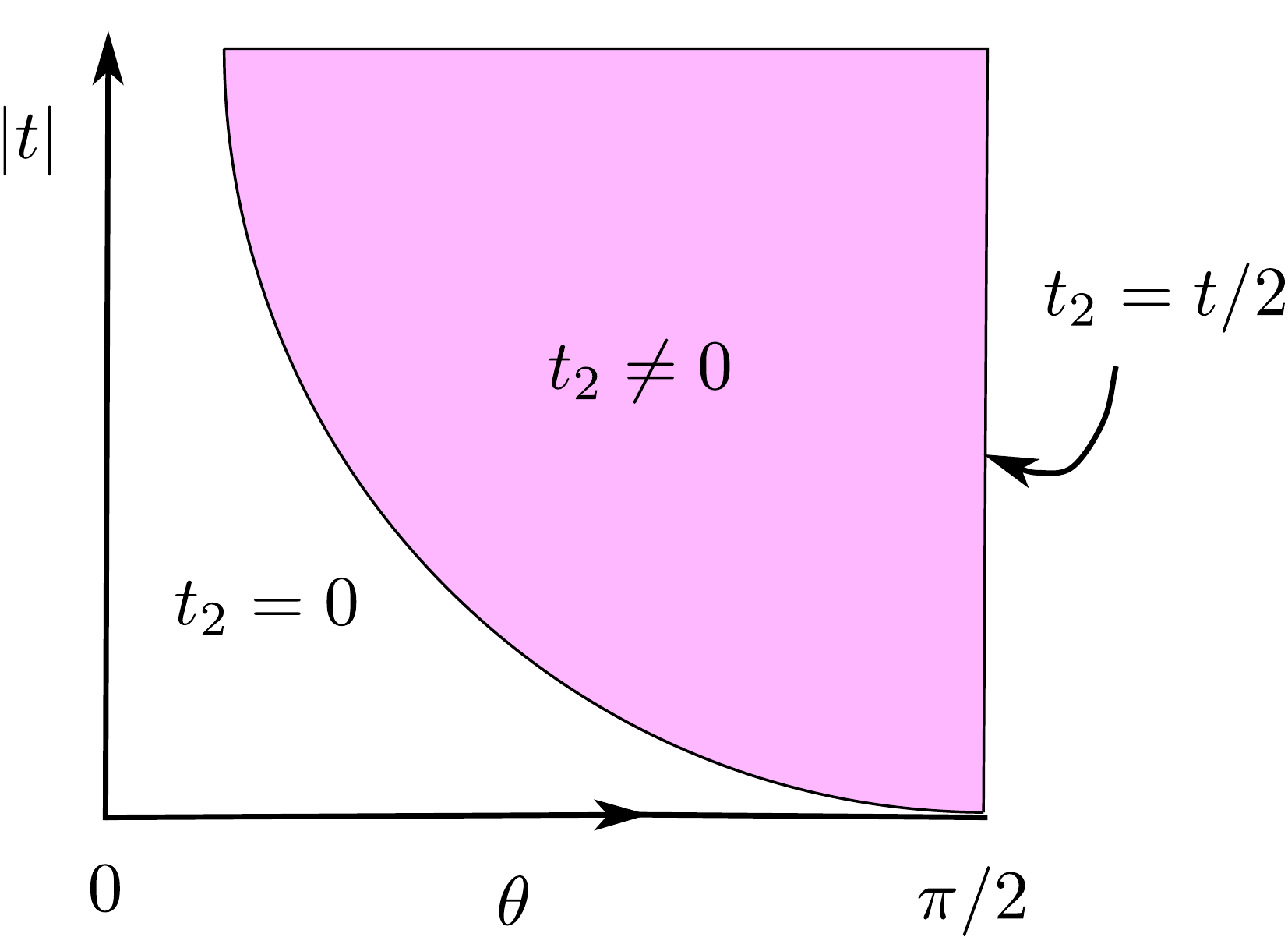}
\end{center}
\caption{The phase diagram describing the large $N$ limit of Chern--Simons theory on $L(2,1)$ for $0\le\theta \le \pi/2$, where $\theta={\rm Arg}(g_s)$. 
There are two phases distinguished by the equilibrium value of $t_2$ and separated by the critical curve (\ref{phaseboundary}). In the 
region $t_2=0$, the $1/N$ asympotics is given by a genus expansion, and corrections due to neighbouring geometries are exponentially suppressed. In the region 
where $t_2\not=0$, neighboring geometries correct the genus expansion with oscillatory terms. Notice that, in this region, the equilibrium value of $t_2$ changes as we change $t$. 
The line $\theta=\pi/2$, where $t_2=t/2$, 
can be regarded as an anti--Stokes line.}
\label{phasediagram}
\end{figure}

The qualitative features of the phase diagram for $0<{\rm Arg}(g_s)<\pi/2$ can be understood by looking at the function (\ref{ainst}). Let us denote
\be
\theta={\rm Arg}(g_s)
\ee
so that
\be
t=|t|\re^{\ri \theta}. 
\ee
If ${\rm Re}(A(t)/g_s)>0$, when $t$, $g_s$ are now complex, the dominant saddle will be $t_1=t$, $t_2=0$, as for $\theta=0$. For $\pi/2>\theta>0$, 
the function
\be
\label{ftheta}
f_{\theta}(|t|)={\rm Re}\left( \re^{-\ri \theta} A\left( \re^{\ri \theta} |t| \right) \right)
\ee
vanishes for a certain critical value 
\be
\label{phaseboundary}
|t|=|t|_{\rm c}(\theta),
\ee
 which can be easily determined numerically. The curve (\ref{phaseboundary}) separates two different regions in the complex $t$ plane, as shown in \figref{phasediagram}. In the region under this curve, where
\be
0\le |t|< |t|_{\rm c}(\theta),
\ee
the dominant saddle is $t_1=t$, $t_2=0$. In the region above this curve, we expect the dominant saddle to have $t_2\not=0$. 

\begin{figure}[!ht]
\leavevmode
\begin{center}
\includegraphics[height=5cm]{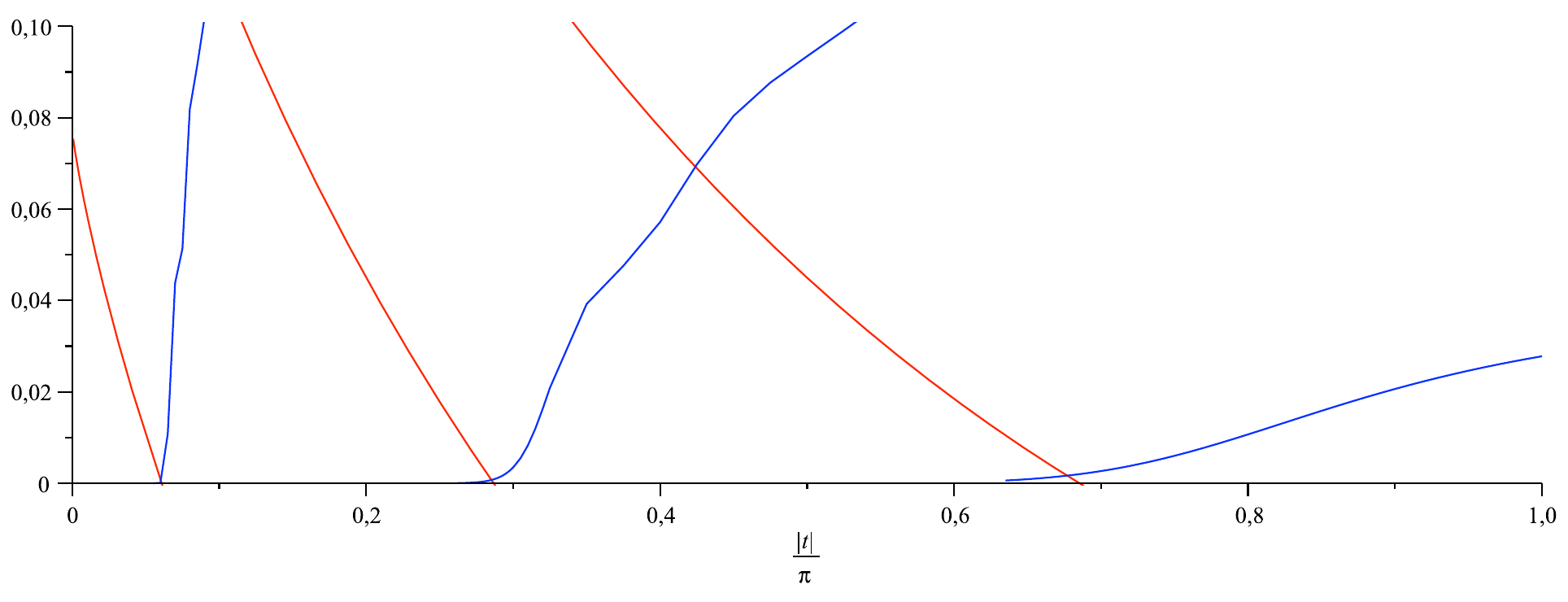}
\end{center}
\caption{Plot of (\ref{n2av}) (blue), calculated for $N=22$, 
and the real part of the instanton action (\ref{ftheta}) (red), as a function of $|t|/\pi$, for $\theta=9\pi/20$, $7\pi/20$ and $\pi/4$, from left to right. The function (\ref{ftheta}) decreases monotonically until it becomes zero at $|t|_c(\theta)$. At this point, the average (\ref{n2av}) gets triggered and it increases with $|t|$. }
\label{thth}
\end{figure}

\begin{figure}[!ht]
\leavevmode
\begin{center}
\includegraphics[height=5cm]{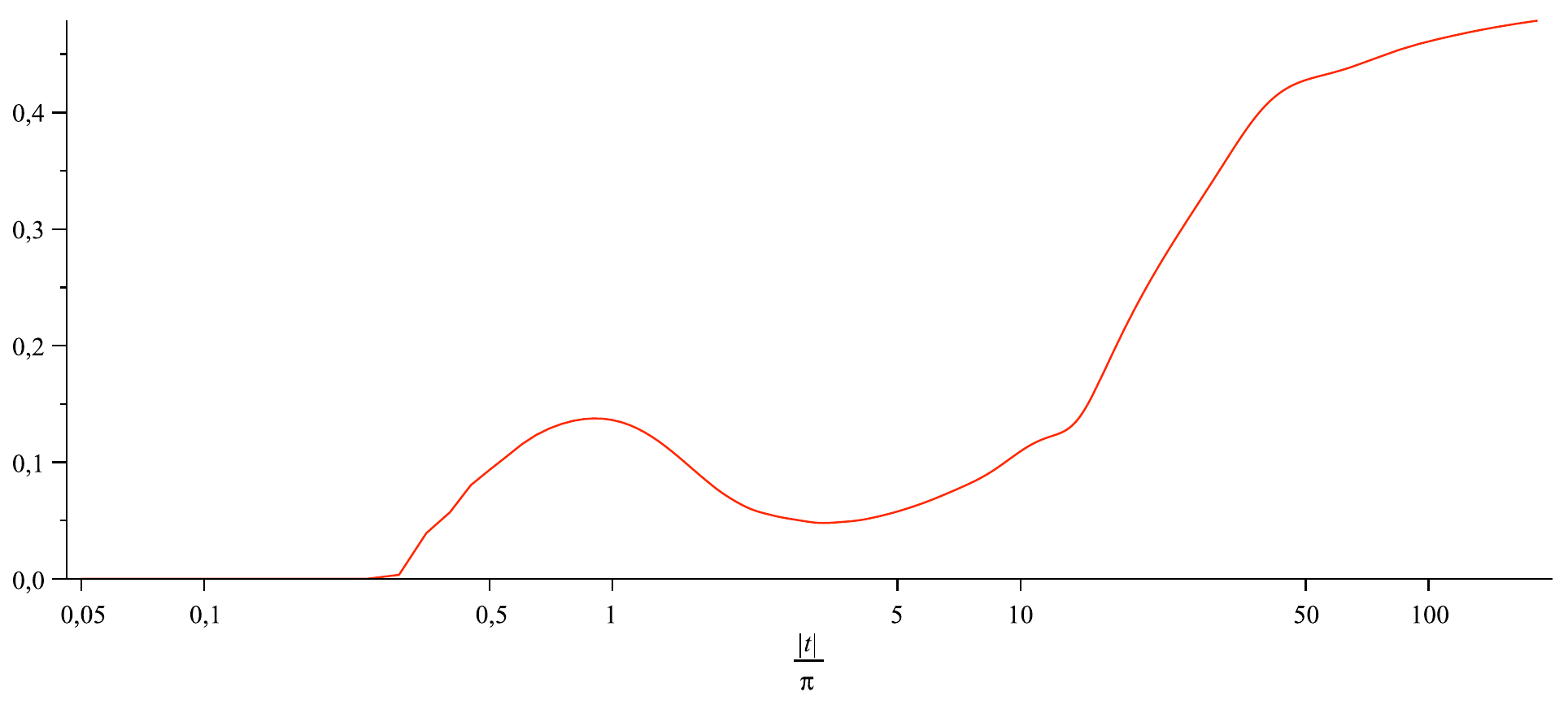}
\end{center}
\caption{Plot of (\ref{n2av}) as a function of $|t|/\pi$, for $N=22$ and $\theta=7\pi/20$. }
\label{complexn2}
\end{figure}

In order to verify this, as well as the location of the phase boundary, we have computed numerically the average filling fraction at finite $N$
\be
\label{n2av}
\left \langle {N_2 \over N} \right \rangle_{\theta} ={1\over  \left| Z(N,g_s) \right|} \sum_{N_2=0}^N {N_2 \over N} \left| Z(N, N_2, g_s) \right|
\ee
for fixed $\theta$, as a function of $|t|$. This gives a finite $N$ approximation to the value of $|t_2/t|$ at the dominant saddle. In \figref{thth} we plot the value of (\ref{n2av}) and the real part of the instanton action (\ref{ftheta}), for three different values of $\theta$, and as a function of $|t|/\pi$. Within the limits of the numerical approximation, we clearly see that $N_2/N$ starts developing an expectation value precisely when  we reach (\ref{phaseboundary}). As shown in \figref{complexn2}, the limiting value of $|t_2/t|$ as 
$|t|$ increases seems to be $1/2$.  

\subsection{Stokes phenomenon and target geometries}

\begin{figure}[!ht]
\leavevmode
\begin{center}
\includegraphics[height=7cm]{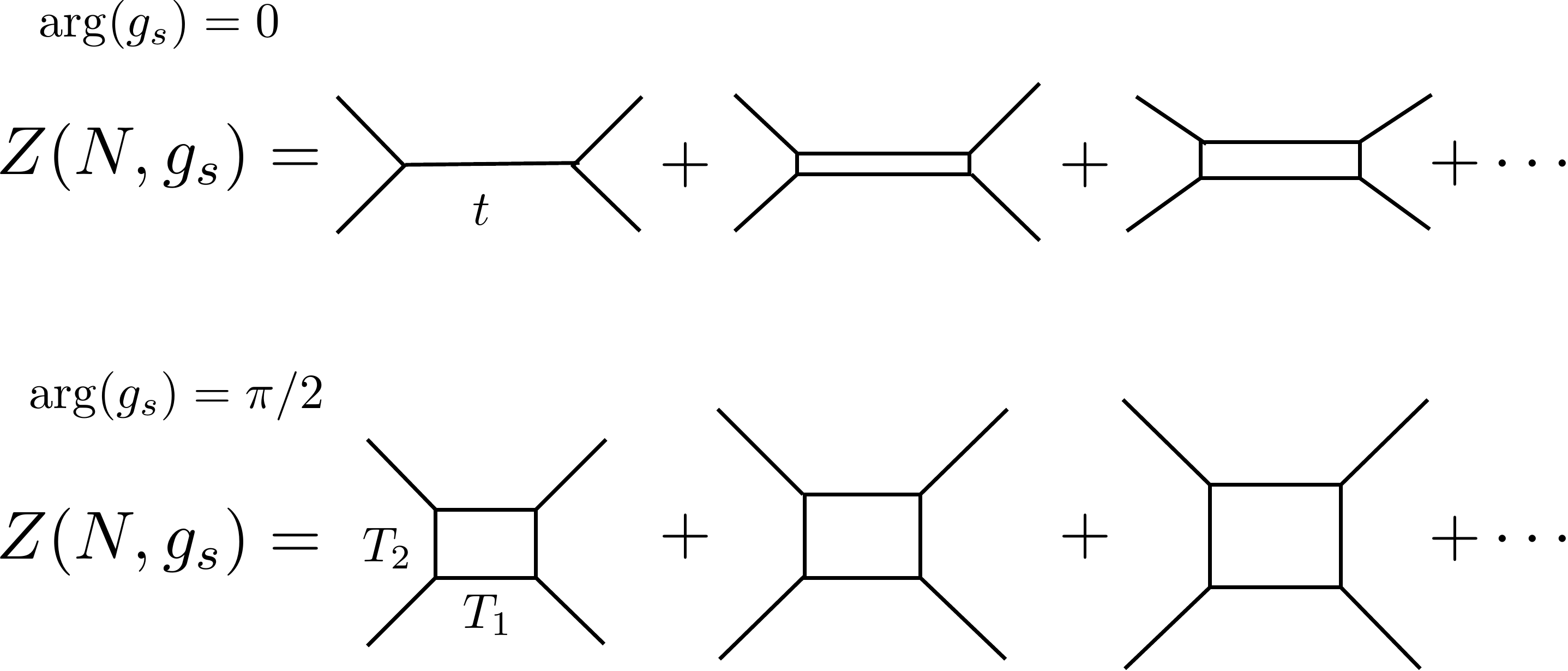}
\end{center}
\caption{The Stokes phenomenon for the large $N$ duality of $L(2,1)$. When $g_s$ is real and positive, the saddle geometry giving the dominating contribution 
to the non-perturbative perturbative function is simply the resolved conifold. The corrections, which are exponentially suppressed, can be interpreted as background geometries 
in which $t_2=\ell g_s$ is quantized in units of $g_s$. 
When $g_s$ is purely imaginary, the saddle geometry is a local $\IP^1 \times \IP^1$ with $t_1=t_2$. Corrections to the genus expansion 
contribute to the large $N$ asymptotics. They can be regarded as a sum of infinite geometries where $t_2=ng_s$ is quantized, and their contribution adds up to a theta function. In terms of the large radius K\"ahler parameters, these geometries have fixed $T_1-T_2=t$. }
\label{saddlegeoms}
\end{figure}

We can now summarize the most important qualitative results of the analysis of the phase diagram. 

First of all, for each complex value of $t$, the large $N$ asymptotics of the non-perturbative free energy is dominated by a fixed filling fraction or background $(t_1^*, t_2^*)$. 
Such a background corresponds to a fixed target geometry, i.e. to a point in the Calabi--Yau moduli space. However, the dominant background changes as we vary $t$. 
 
The second important point is that the structure of the large $N$ asymptotics depend on the nature of the dominant 
saddle point. If the dominant saddle point is on the boundary, i.e. one of the filling fractions 
vanishes, the $1/N$ asymptotics is given by a conventional genus expansion. There are non-perturbative effects due to the contributions of 
neighbouring geometries, as shown in (\ref{npboundary}), but they are exponentially suppressed. However, if the dominant saddle point is an interior point, 
the asymptotics has an oscillatory behavior and it involves theta functions, as detailed in (\ref{npinterior}). In particular, the large $N$ asymptotics is no longer given by a genus expansion around a fixed geometry, and corrections due to neighboring geometries are crucial already at the next-to-leading order. We hope that our detailed numerical analysis in the case 
$\theta=\pi/2$ has convinced the reader that these effects are crucial in order to obtain the correct large $N$ asymptotics of the gauge theory partition function. 

It is interesting to notice that the non-perturbative corrections are due to fluctuating target geometries whose size is {\it quantized} in units of $g_s$,
see Fig. \ref{saddlegeoms}. In the case of the boundary expansion, an $\ell$-instanton correction corresponds to a geometry with $t_2=\ell g_s$, $\ell=1, 2, \cdots$. In the interior case, the theta function 
is a discrete sum over an infinite number of geometries with $t_2=(n-N/2)g_s$, $n\in \IZ$. Notice that the saddle geometries for ${\rm arg}(g_s)=\pi/2$ correspond to the region in moduli 
space where $z_1, z_2\rightarrow \infty$ with $z_1/z_2$ fixed and finite, therefore they are not in the large radius phase $z_1, z_2 \rightarrow 0$.

Finally, we point out that the change of saddle geometry as we change the complex 't Hooft parameter can be regarded as a {\it generalized Stokes phenomenon}. 
In fact, in the limit of vanishing 't Hooft coupling, 
the asymptotics changes {\it discontinuously} as we change the argument of the string coupling constant: we have a sudden jump from a dominant saddle at $t_2=0$ to a dominant saddle at $t_2/t=1/2$. The jump in the asymptotics has the same origin as in the Stokes phenomenon: saddles which were subleading due to ${\rm Re}(A/g_s) >0$ are no longer 
suppressed exponentially, and they lead to an oscillatory asymptotics in the $\theta =\pi/2$ direction. In fact, the line $\theta=\pi/2$ plays the role of an anti--Stokes line. 
These analogies can be made more precise by studying a model where the connection to the standard Stokes phenomenon is manifest and can be followed in detail, namely the 
cubic matrix model. We will show in the next section that the phase diagram for this model is very similar to the one we have just found, and we will able to confirm all the analogies 
that we have stated. 

It is important to notice that 
the connection to the standard Stokes phenomenon is only strictly true when $t\rightarrow 0$. For finite $|t|$ the Stokes phenomenon is {\it smoothed out}, in the sense that 
for fixed $|t|$, as we increase $\theta$ from $\theta =0$, 
the asymptotics changes smoothly from a phase with $t_2=0$ to a phase with $t_2 \not=0$. The phase boundary (\ref{phaseboundary}) 
corresponds then to a second-order phase transition with $t_2$ as its order parameter.

\subsection{Background independence}

Our results on the phase structure of this model shed some light on the issue of background independence in topological string theory. 
As it is well-known, the genus $g$ free energies in topological string theory, $F_g(t_i)$, are quasi-modular forms under the action of the 
symplectic group \cite{abk}. This can be regarded as a consequence of the holomorphic 
anomaly of \cite{bcov}, and it means in particular that the $F_g(t_i)$ are not well-defined functions on the moduli space.      

In the models we are studying, thanks to the large $N$ duality, there is a quantity which is well-defined non-perturbatively, namely the 
Chern--Simons theory partition function, which depends on $g_s$ and the rank $N$ of the gauge group. Notice that, as needed for consistency, the total 
't Hooft parameter $t$ is a symplectic invariant in the dual topological string. 
An important question is how this non-perturbative 
quantity is related to the perturbative, background dependent topological string free energies. In this paper we have provided a detailed answer to this question in the 
particular example of $L(2,1)$ and its string dual. 

First of all, at finite $N$ the non-perturbative partition function (\ref{ccp}) is a {\it sum} over all 
possible backgrounds $N_1+N_2=N$. In the gauge theory description, these backgrounds correspond to saddle points or flat connections. In the 
matrix model description, they correspond to different filling fractions. In the language of statistical mechanics, we can say 
that the background-independent, non-perturbative partition function has to be calculated in a grand-canonical ensemble where $N_2$ is a fluctuating variable. This is manifest in (\ref{ccp}), where the parameters $\zeta_i$ play the role of fugacities. On the other hand, the background-dependent 
topological string partition function is computed in a canonical ensemble where $N_2$ is fixed. 

In the thermodynamic limit (i.e. at large $N$), as it is well-known in statistical mechanics, the sum over backgrounds is strongly peaked around a particular one, and both 
ensembles are equivalent. This particular background is the 
dominant saddle which we have determined in various situations. It depends on the gauge theory parameters $N$ and $g_s$, and it changes 
as we move on the complex $t$ plane, displaying the rich phase structure that we have analyzed. At finite $N$, however, the results of the two ensembles 
are different, and they differ precisely in the terms which go beyond the genus expansion, i.e. in the contributions of large $N$ multi-instantons. 

The resulting non-perturbative picture is very different 
from the picture that one obtains in perturbative topological string theory: the background-dependent quantities are {\it emergent} quantities and they only make 
sense in the large $N$ limit, since they are defined through 
an asymptotic expansion.

\sectiono{Stokes phenomenon and matrix models}

\begin{figure}[!ht]
\leavevmode
\begin{center}
\includegraphics[height=7cm]{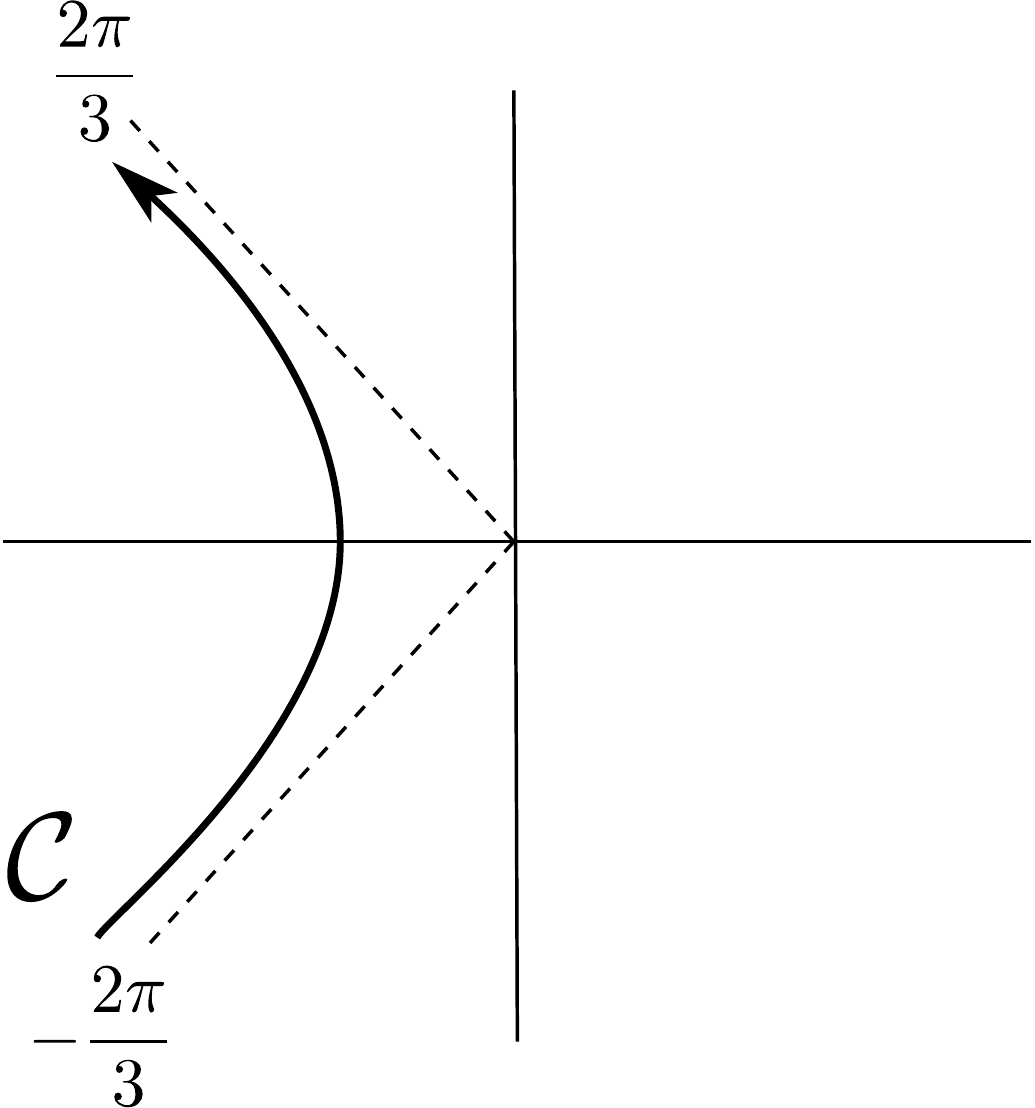}
\end{center}
\caption{The integration contour $\CC$ for (\ref{zoneg}), which leads to a convergent integral.}
\label{integrationpath}
\end{figure}

In order to have a better understanding of the nature of the line ${\rm Arg}(g_s)=\pi/2$ in the phase diagram of Chern--Simons theory, we will make a detailed analysis of the 
Stokes phenomenon in a simple matrix model. We will focus on the quintessential example of Stokes phenomenon, namely the Airy function, and we will study its fate as we promote the Airy integral to a matrix integral. Of course, the study of matrix models in the double-scaling limit produces differential equations which display the 
so-called non-linear Stokes phenomenon \cite{moore,painlevebook}. Here we are rather interested in recovering the standard, {\it linear} Stokes phenomenon in the 
context of matrix integrals. 

\subsection{Review of the Stokes phenomenon for the Airy function}

\begin{figure}[!ht]
\leavevmode
\begin{center}
\includegraphics[height=7cm]{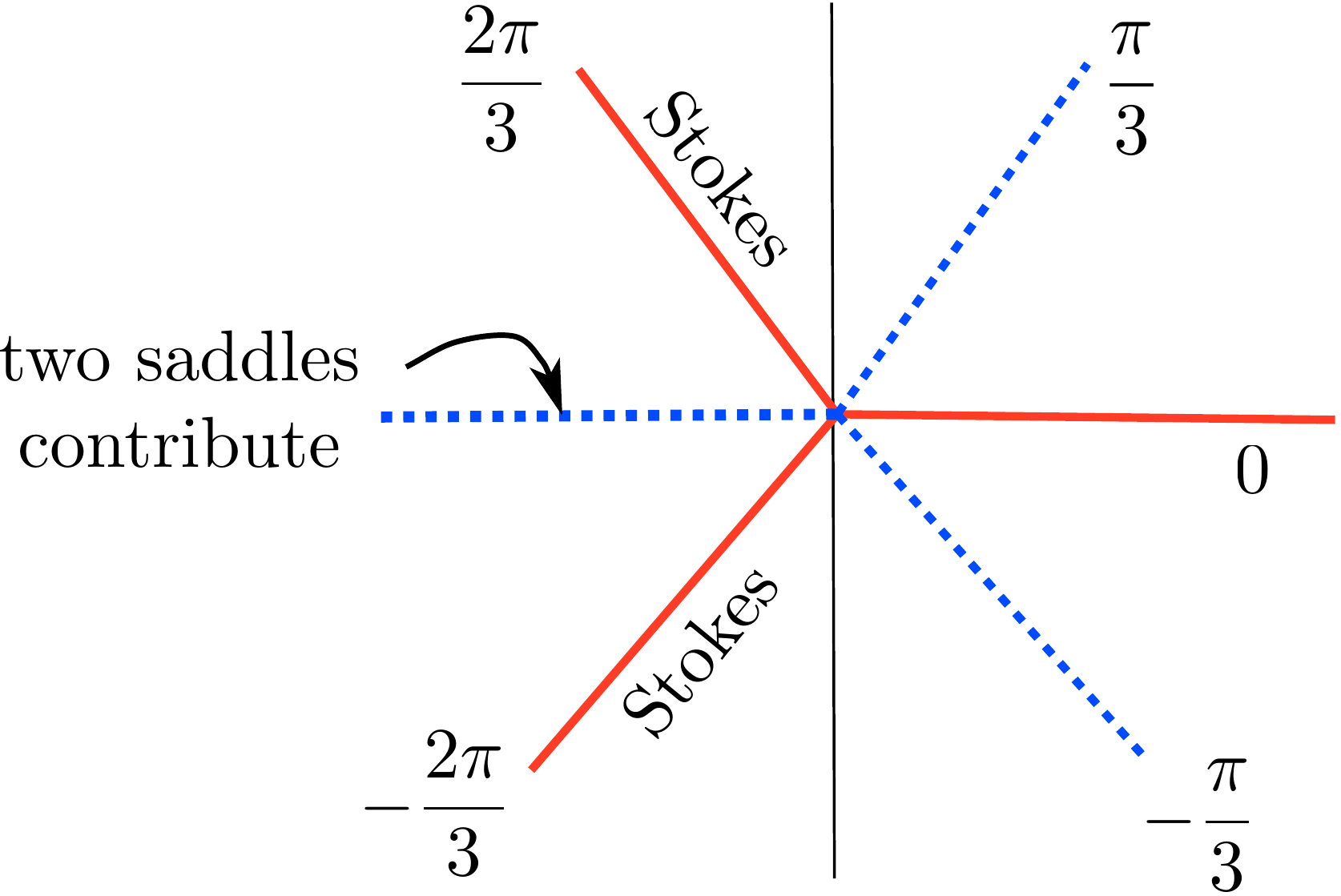}
\end{center}
\caption{Saddle-point analysis of the asymptotics of the function (\ref{zoneg}), closely related to the Airy function ${\rm Ai}(\zeta)$. Full lines (in red) are Stokes lines, while 
dashed lines (in blue) are anti-Stokes lines. On the Stokes lines $\kappa =\pm 2\pi/3$, a second saddle appears in the 
integration contour. This saddle is subdominant when $2\pi/3\le |\kappa|<\pi$ and does not contribute to classical asymptotics. However, at $\kappa=\pi$, the saddle 
is not subdominant anymore and leads to an oscillatory asymptotics.}
\label{airystokes}
\end{figure}

Let us consider the potential 
\be
\label{stokespot}
 V(x)= -\re^{\ri \kappa} x + {x^3 \over 3}
\ee
where $\kappa \in [0,2\pi]$. The one-dimensional integral
\be
\label{zoneg}
Z_1(g_s)=\int_{\CC} {\rd x \over 2\pi} \re^{-{1\over g_s} V(x)}
\ee
where $\CC$ is the path shown in \figref{integrationpath}, and $g_s$ is real and positive, is essentially the Airy function ${\rm Ai}(w)$, where
\be
\label{vzeta}
w=g_s^{-2/3} \re^{\ri \kappa}.
\ee
The small $g_s$ asymptotics of this integral is given by the large $w$ asymptotics of the Airy function. The standard analysis of this asymptotics is 
as follows (see \cite{miller} for a very nice discussion). There are two saddle points 
\be
x^{{\rm L}, {\rm R}}=\mp \zeta^{1/2}
\ee
where we have introduced the variable
\be
\zeta=\re^{\ri \kappa}.
\ee
These saddles have actions
\be
\label{sadaction}
-{1\over g_s} V(x^{{\rm L}, {\rm R}})=\mp {2\over 3 g_s} \zeta^{3/2}.
\ee
For 
\be
 - {2\pi \over 3}< \kappa < {2\pi \over 3}
\ee
the path $\CC$ can be deformed into a path of steepest descent through the saddle point at $x^{\rm L}$. When $\kappa=2\pi/3$,
the steepest descent path coming from the saddle at $x^{\rm L}$ runs right into the other saddle point. At this angle we have
\be
{\rm Im}(V(x^{\rm L}))={\rm Im}(V(x^{\rm R}))
\ee
and the corresponding direction is called a Stokes line. This is the place where the second saddle $x^{\rm R}$ might start contributing to the integral. In fact, for 
\be
{2\pi \over 3}< |\kappa|<\pi
\ee
the contour $\CC$ gets deformed into a steepest descent path passing through $x^{\rm L}$ {\it together} with a steepest descent path 
passing through $x^{\rm R}$. However, in this range the latter gives an exponentially suppressed contribution, and the classical 
asymptotics is just given by the contribution from $x^{\rm L}$. This leads to 
\be
Z_1 \sim \re^{-2w^{3/2}/3}
\ee
where $w$ is the variable introduced in (\ref{vzeta}). When $\kappa=\pi$ the asymptotics is different, since both saddles have the same real part
\be
{\rm Re}(V(x^{\rm L}))={\rm Re}(V(x^{\rm R})).
\ee
A line where this occurs is called an anti-Stokes line. Along this line both saddles contribute to the asymptotics, and we have
\be
Z_1 \sim \cos \Bigl( {2\over 3} |w|^{3/2} -{\pi \over 4}\Bigr).
\ee
The fact that different asymptotic formulae hold on different directions for the same analytic function is called the Stokes phenomenon. From the point of saddle-point analysis, 
what is happening is that the saddle point which appeared on the Stokes lines, at $\kappa=\pm 2\pi/3$, is no longer subdominant at $\kappa=\pi$, and it has to be included in the asymptotics. The saddle-point analysis is summarized in \figref{airystokes}. We should mention that it is important sometimes 
to take into account the subleading exponentials, even in the region $\kappa \not=\pi$. 
This is very clearly shown in the analysis of Berry in \cite{hyper}, and it has led to the development of 
non-classical asymptotic theories. The most sophisticated of these, resurgent analysis, is applied to the Airy function in for example \cite{delabaere}. 
 
\subsection{The Stokes phenomenon and the cubic matrix model}
Let us now promote (\ref{zoneg}) to a full matrix integral, 
\be
\label{cubicmm}
Z^{\CC}_{\rm np}(N, g_s, \kappa)=\int_{\CC} \prod_{i=1}^N {\rd x_i \over 2\pi} \, \Delta^2(x) \, \re^{-{1\over g_s} \sum_{i=1}^N V(x_i)},
\ee
where all the eigenvalues are integrated along the contour $\CC$. We can now study its small $g_s$ asymptotics for $t=g_s N$ fixed. Of course, this 
is nothing but the 't Hooft expansion of this matrix model. The general saddle point is a {\it two-cut configuration}, labelled by $(N_1, N_2)$, where $N_1$, $N_2$ are the number of eigenvalues near the critical points $x^{{\rm L}, {\rm R}}$, respectively. Before starting the analysis, 
let us notice that for small $t$ the saddle point structure should be the same 
as for the Airy function. The reason is that, for small $t$, the Vandermonde repulsion among eigenvalues is suppressed w.r.t. the potential, and the model 
becomes just $N$ copies of the one-dimensional integral. 

Therefore, at least for small $t$, we expect the following phase structure. For $|\kappa|<2 \pi/3$, the dominant saddle is a one-cut configuration
where all the 
eigenvalues sit near $x^{\rm L}=-\zeta^{1/2}$. This is the boundary saddle point $(N, 0)$. 
It is a one-cut configuration which can be analyzed with standard techniques. The endpoints of the cut $(a,b)$ are determined by the equations
\be
x_0(x_0^2-\zeta)=t, \qquad \delta^2=2(\zeta-x_0^2)\ee
where
\be
a=-x_0+\delta, \qquad b=-x_0-\delta
\ee
and we choose the root $x_0=\zeta^{1/2} +\cdots$. 
The spectral curve is 
\be
y(x) =(x-x_0) {\sqrt{x^2+2xx_0 + 3x_0^2-2\zeta}}
\ee
and the effective potential reads
\be
\ba
V_{\rm eff}(x)&=\frac{1}{3}\left( x (x -x_0) -2 \zeta\right) {\sqrt{x^2+2xx_0 + 3x_0^2-2\zeta}} \\
&-2 x_0 \left(x_0^2-\zeta \right) \log \left(x_0+x+{\sqrt{x^2+2xx_0 + 3x_0^2-2\zeta}}\right).
\ea
  \ee
 For $2\pi/3\le |\kappa|<\pi$, the second saddle should start contributing, but at least at small $t$ it should be exponentially suppressed. In other words, the 
 saddle configuration for the matrix integral in this region should still be the boundary saddle $(N,0)$, but there will be corrections of the form (\ref{npboundary}). The instanton action is given by
   \be
   A=V_{\rm eff}(x_0)-V_{\rm eff}(b).
   \ee
The small $t$ expansion of this action is 
\be
A=-{4\over 3} \zeta^{3/2}-t + t \log t -3 t \log (2 \zeta^{1/2})
-{7 \over 8 \zeta^{3/2}} t^2 +{41  \over 64 \zeta^3} t^3 +\cdots
\ee
   Notice that, as $t\rightarrow 0$, we find
   \be
   x_0 \rightarrow x^{\rm R}, 
   \ee
   which is the other saddle, and
   \be
   A\rightarrow -{4\over 3}\zeta^{3/2}
   \ee
  which is the difference between the actions of the two saddles (\ref{sadaction}). This is in agreement with the expectation that for small $t$ the saddle-point structure 
  of the matrix model is just the one coming from the Airy function. The real part of the instanton action vanishes at a critical value of $t_c(\kappa)$, which depends on the 
  value of $\kappa$. For example, for $\kappa=2\pi/3$, we have
   \be
   t_c= {2\over 3 {\sqrt{3}}}.
   \ee
This is precisely the critical point leading to pure 2d gravity. For $t>t_c(\kappa)$, a phase transition occurs to a new phase, in which we expect generically $N_2^*\not=0$. 

\begin{figure}[!ht]
\leavevmode
\begin{center}
\includegraphics[height=7cm]{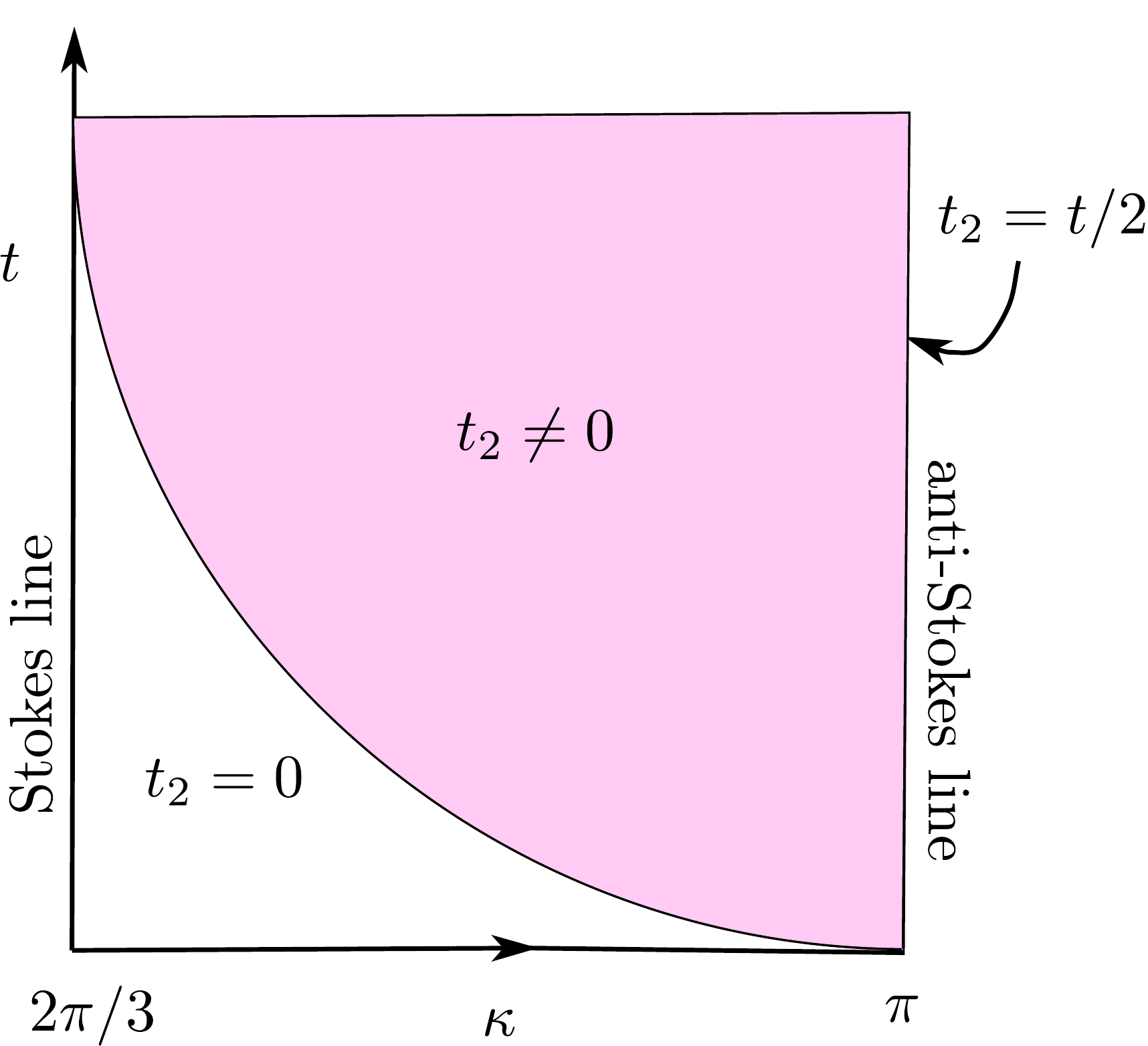}
\end{center}
\caption{The phase diagram of the cubic matrix model (\ref{cubicmm}) as a function of $2\pi/3\le \kappa\le \pi$ and $t=g_s N$ (which is taken to be real), for small $t$. 
The Stokes and anti--Stokes lines of the $N=1$ case, which occur for $\kappa=2\pi/3$ and $\kappa=\pi$, respectively, determine to a large extent the 
phase diagram for small $t$. On the anti--Stokes line, the saddle value of $t_2$ is $t/2$, and the classical $1/N$ asymptotics requires corrections beyond the genus expansion. }
\label{cubicphase}
\end{figure} 
  
When $\kappa=\pi$, i.e. on the anti--Stokes line, 
we expect that the standard Stokes phenomenon controls the behaviour of the matrix integral for small $t$. Indeed, this is the case. The real part of the 
action $A(t)$ vanishes at $t=0$, and it is actually negative for small $t$. We then have to look for a new saddle point by analyzing the behavior of $F_0'$. The genus zero free energy 
of the cubic matrix model has been analyzed in many places \cite{civ,kmt}. It has the structure 
\be
F_0(t_1, t_2) = F^{\rm np}_0(t_1, t_2)+ F^{\rm p}_0(t_1, t_2), 
\ee
where
\be
F^{\rm np}_0(t_1, t_2)=
{1\over2}t_1^2 \Bigl( \ln \, t_1-{3\over 2}\Bigr)+
{1\over2}t_2^2 \Bigl( \ln \, t_2-{3\over 2}\Bigr)-{1\over 2} (t_1^2 + t_2^2) \log(2 \zeta^{1/2}), 
\ee
and
\be
\label{perfo}
\ba
F^{\rm p}_0(t_1, t_2) &= -{2\over 3} \zeta^{3/2} (t_1- t_2) + 2t_1 t_2 \log(2 \zeta^{1/2})+{1\over 8\zeta^{3/2}}\Bigl(-{2\over3}t_1^3-5 t_2^2 t_1+5 t_2 t_1^2
+{2\over3}t_2^3\Bigr)\\
&+{1\over 64 \zeta^3}\Bigl({8\over3}t_1^4-{91\over3}t_1^3 t_2+59 t_1^2 t_2^2
-{91\over3}t_1 t_2^3+{8\over3}t_2^4\Bigr)+\cdots
\ea
\ee
On the anti--Stokes line ${\rm arg}(\kappa)=\pi$, $\zeta^{1/2}$ is purely imaginary, and it is easy to see from the structure of the genus zero free energy that its {\it real} part is {\it symmetric} in $t_1$, $t_2$. Therefore, 
\be\label{cubicts}
t_1=t_2
\ee
solves the saddle-point equation 
\be
{\rm Re}\, {\partial F_0 \over \partial s}=0
\ee
at least for small $t$ (more precisely, (\ref{cubicts}) provides a solution in the domain of analiticity of the convergent expansion (\ref{perfo})). Therefore, on the anti--Stokes line and for 
$t$ small enough, the partition function of the cubic matrix model (\ref{cubicmm}) is given by an expansion of the form (\ref{npinterior}). This type of expansion is then the natural generalization 
to matrix models of the oscillatory behavior along an anti--Stokes line.  

The phase diagram of the cubic matrix model, again at small $t$, is represented in \figref{cubicphase}. 
We see that it is very similar to the phase diagram of the large $N$ Chern--Simons theory on a 
lens space that we analyzed in the previous section. This confirms that the line $\theta=\pi/2$, where the vacuum occurs at $t_1=t_2=t/2$, plays the role of an anti--Stokes line. 
Strictly speaking, however, the discontinuous jump in the asymptotics typical of the Stokes phenomenon occurs only at $t=0$. For $t\not=0$, as we change the angle we encounter the 
region where $t_2$ goes from a zero value to a nonzero value in a smooth way. We conclude that turning on the 't Hooft parameter smooths out the Stokes discontinuity, and we pass 
from a first-order phase transition at $t=0$ to a second-order phase transition for $t>0$, where $t_2$ plays the role of an order parameter. 

As $t$ increases along the different directions, further phase transitions can occur which we have not explored. For example, along the $\kappa=\pi$ direction, it follows from the results of \cite{bertola} that there is a transition at $t>0$ in which the two cuts merge into a trivalent graph. Interestingly, such transitions 
along the anti--Stokes line are absent in the lens space partition function. 
 
\sectiono{Conclusions}

In this paper we have analyzed non-perturbative aspects of the large $N$ expansion in a theory with a rich structure of instanton sectors and with a large $N$ dual. 
There are various important conclusions that emerge from our analysis. First of all, we have seen that, for each choice of gauge theory parameters $g_s$, $N$, the large $N$ 
asymptotics is dominated by a single instanton sector. This sector corresponds to a fixed geometry in the moduli space of the dual Calabi--Yau geometry, which can be characterized by the Boutroux condition (\ref{boutrouxc}). Second, there are large $N$ phase transitions as we move in the complex $t$ plane, and different instanton sectors dominate the asymptotics 
for different values of $t$. These transitions can be regarded as deformations of the standard Stokes phenomenon. 
Third, the correct large $N$ asymptotics goes beyond the genus expansion, and subleading saddles must be incorporated. 



We conclude with a list of issues left open by our work:
\begin{enumerate}
\item 
The genus $g$ topological string free energies appear in the large $N$ asymptotics of the ``canonical" partition function $Z(N, N_2, g_s)$ 
with $N_2$ fixed. They are known to transform as quasi-modular forms when moving from one patch of the moduli space to another \cite{abk}.
At the level of the gauge theory, these transformations should correspond to phase transitions within the same instanton sector,
which we have not studied in detail in this paper, since we have focused on phase transitions among different instanton sectors. 
On the other hand, the matrix integral (\ref{fixedn2}) gives a non-perturbative definition of this ``canonical" partition function which is valid for any $N_k$ and any complex $g_s$. 
It would be important to understand the precise connection between the modular properties of the topological string amplitudes and the large $N$ asympotics of the quantity (\ref{fixedn2}),
which is background independent and clearly well-defined in all the moduli space. The results obtained in this paper seem to indicate that modularity can be understood 
in terms of phase transitions or generalized Stokes phenomena, and that background dependence in these models is ultimately an artifact of the asymptotic large $N$ expansion.

\item  Non-perturbative effects of the form $\re^{-A/g_s}$ in gauge theories and matrix models have been identified in terms of D-branes in their string duals.
In particular, the source of these effects has been shown to be 
Liouville branes in minimal strings, and more recently toric branes in topological strings \cite{open,mswone,ps}. 
It seems clear that the subleading saddles appearing in the oscillatory asymptotics (\ref{npinterior}) can be also interpreted in terms 
of D-branes, and in the example of local $\IF_0$ these are also toric D-branes in this geometry. 
It would be very interesting to make more precise the D-brane interpretation of the non-perturbative corrections appearing in (\ref{npinterior}).

\item Even though in this paper we focused on a simple toy model of gauge/string duality, we believe that some of the effects we 
studied in this work are likely to play a role 
in other large $N$ dualities. The underlying reason for the phenomena explored in this paper is the 
existence of nontrivial instanton sectors which compete among them at large $N$. In some cases we have a single sector dominating the large $N$ asymptotics, 
but in other cases, along generalized anti--Stokes lines, the standard large $N$ asymptotics needs to be corrected, as first pointed out in \cite{bde}. This scenario seems to be quite 
general and it is likely to appear in other models, in particular in AdS/CFT dualities. For example, the recent work \cite{kapustin} indicates that matrix integrals closely related to 
(\ref{intdef}) are relevant in the study of large $N$ dualities for ABJM theories.

\end{enumerate}

\section*{Acknowledgments}
We would like to thank Bertrand Eynard, Andreas Malaspinas and Cumrun Vafa for discussions and suggestions. The work of M.M. and P.P. is supported in part by the 
Fonds National Suisse. 


 \appendix

\sectiono{Some aspects of topological string theory on the local $\IP^1\times \IP^1$ geometry}

In this Appendix we give some further information on topological string theory on the local $\IP^1\times \IP^1$ geometry, and in particular we justify some of the statements made in the bulk of the text. 

The Yukawa couplings of this Calabi--Yau, when written in terms of the ``bare" coordinates $z_1, z_2$ appearing in (\ref{mc}), are \cite{akmv,hkr}
 \be
 \ba\label{yuka}
 C_{111}&={(1 - 4 z_2)^2 - 16 z_1 (1 + z_1) \over 4 z_1^3 \Delta}, \\
 C_{112}&={16 z_1^2 - (1 - 4 z_2)^2 \over 4 z_1^2 z_2 \Delta},\\
 C_{122}&= {16 z_2^2 - (1 - 4 z_1)^2 \over 4 z_1 z_2^2 \Delta},\\
 C_{222}&={(1 - 4 z_1)^2 - 16 z_2 (1 + z_2) \over 4 z_2^3 \Delta},
 \ea
 \ee
where
\be
\Delta=1-8(z_1+z_2)+16 (z_1-z_2)^2.
\ee

Using the Yukawa couplings (\ref{yuka}), it is possible to find a closed form for
the prepotential when $x_2=0$. The triple derivative w.r.t $x_1$ is given by
\be
C_{x_1x_1x_1}\Big |_{x_2=0}=\left(C_{111}J_1^3+3C_{112}J_1^2J_2+3C_{122}J_1 J_2^2+C_{222}J_2^3\right)_{x_2=0}=\frac{2-x_1}{4x_1(x_1-1)^3},
\ee
where
\be
J_1=\frac{\partial z_1}{\partial x_1}=\frac{x_1-2}{x_1^3 x_2^2},\ \ J_2=\frac{\partial z_2}{\partial x_1}=-\frac{2}{x_1^3 x_2^2}.
\ee 
Since
\be
\partial_{x_1} \sigma_1=\frac{1}{1-x_1},
\ee
 we obtain a closed form for the triple derivative w.r.t. the  flat coordinate $\sigma_1$:
 \be
 C_{\sigma_1\sigma_1\sigma_1}\Big |_{x_2=0}=\frac{1+\re^{-\sigma_1}}{4(1-\re^{-\sigma_1})},
 \ee
 which can be integrated to give:
\be
\mathcal{F}_0(\sigma_1,0)=-\frac{1}{2} {\rm Li}_3(\re^{-\sigma_1})+\cdots
\ee
The dots indicate a degree three polynomial. This polynomial can be fixed by comparing to the perturbative results, and one finally obtains (\ref{f0}).

It is also possible to derive closed expressions for $\partial_{\sigma_2} \CF_{\sigma_2}$ on the slice $x_2=0$. To do this, one first writes the Picard--Fuchs system (\ref{pfsystem}) in terms of 
the coordinates $x_{1,2}$, 
\be
\ba
 \mathcal{L}_1&=\frac{1}{4} \left(8-8 x_1+x_1^2\right) x_2 \d_{x_2}+\frac{1}{4} \left(-4+\left(-2+x_1\right){}^2 x_2^2\right) \d^2_{x_2}\\& 
 +\left(-1+x_1\right) x_1^2 \d_{x_1}-x_1 \left(2-3 x_1+x_1^2\right) x_2 \d_{x_1}\d_{x_2}+\left(-1+x_1\right){}^2 x_1^2 \d^2_{x_1},\\
 \mathcal{L}_2&=\left(2-x_1\right) x_2 \d_{x_2}+\left(-1+\left(1-x_1\right) x_2^2\right) \d^2_{x_2}-x_1^2 \d_{x_1}\\ 
 &+2 \left(-1+x_1\right) x_1 x_2 \d_{x_1}\d_{x_2}+\left(1-x_1\right) x_1^2 \d_{x_1}^2\;,
 \ea
 \ee
and looks for a solution $\CL_{1,2}f=0$ in the form
\begin{equation}
 f(x_1,x_2)=x_2G(x_1)+\CO(x_2^3).
 \label{x_2_expansion}
\end{equation}
The PF system leads to the following equation for $G(x)$:
\begin{equation}
 \frac{1}{4}(x-4)\,G(x) + x\,G'(x) - (1 - x)x^2\,G''(x)=0,
 \end{equation} 
whose general solution is a linear combination of $xK(x)$ and $xK(1-x)$. By comparing to the first terms in (\ref{speriods}) and using (\ref{pf0}), one immediately obtains
\begin{equation}
\label{firstseries}
 \left.\frac{\d \sigma_2}{\d x_2}\right|_{x_2=0}=\frac{2}{\pi}x_1K(x_1), \qquad 
{\partial \over \partial x_2} \left( {\partial F_0 \over \partial s} \right)_{x_2=0}=-x_1K(1-x_1),
\end{equation} 
therefore
\begin{equation}
\label{secondseries}
 {\partial^2 F_0 \over \partial s^2}\bigg|_{x_2=0}=
4 \left(\left.\frac{\d \sigma_2}{\d x_2}\right|_{x_2=0}\right)^{-1}{\partial \over \partial x_2} \left( {\partial F_0 \over \partial s} \right)_{x_2=0}=-2 \pi \frac{ K(1-x_1)}{K(x_1)},
\end{equation} 
and (\ref{tau}) follows. Equivalently, one can derive (\ref{firstseries},\ref{secondseries}) by solving the recursion (\ref{cdrecursion}) for $c_{m,1}$ and $d_{m,1}$, 
\be
\ba
c_{m,1}&=\frac{\Gamma(m-1/2)^2}{\pi \Gamma(m)^2},\ \ m\geq 1, \\ 
d_{m,1}&=-2(\psi_{m}-\psi_{m-1/2}) c_{m,1},
\ea
\ee
and using the series expansions 
\be
\ba
K(x)&=\frac{\pi}{2}\sum_{k=0}^\infty \left(\frac{1}{2}\right)_k^2 \frac{x^k}{(k!)^2},\\ 
K(1-x)&=-\frac{1}{\pi} \log(x)K(x)+\sum _{k=0}^{\infty }\left(\frac{1}{2}\right)_k^2 \frac{x^k}{(k!)^2}(\psi_{n+1}-\psi_{n+1/2}).
\ea
\ee

Finally, we calculate the instanton action (\ref{ainst}) in the slice $t_2=0$. This can be in principle derived from the above geometry, but in fact we can 
use a computation already done in \cite{abms}. As in \cite{mswone,mswtwo}, the one--instanton free energy is given by the 
ratio of the one instanton partition function $Z_N^{(1)}\equiv Z(N,N_2=1,g_s)$ 
and the zero instanton partition function $Z_N^{(0)}\equiv Z(N,N_2=0,g_s)$,
\be
\label{ratio}
F^{(1)}=\frac{Z^{(1)}_N}{Z^{(0)}_N}\sim \re^{-A(t)/ g_s},
\ee
We can evaluate $Z^{(1)}_N$ by taking  $(N_1,N_2)=(N-1,1)$ in the matrix integral (\ref{intdef}), as in \cite{mswone}, and we find
\be
Z^{(1)}_N=\re^{-{\pi^2 \over 2 g_s} -g_s (\rho^2_N -\rho^2_{N-1})}Z^{(0)}_{N-1}
 \int  {\rd s \over 2\pi} \, {\rm e}^{-{1\over 2 g_s}  s^2}
 \left \langle \prod_{j=2}^N \Bigl( 2 \sinh {s+\pi \ri - x_j\over 2}
 \Bigr)^2
\right \rangle_{N-1},
\ee
where $\rho_N$ is the Weyl vector in the $U(N)$ theory, and the vev is calculated in the matrix model defined by $Z_N^{(0)}$. 
At small $g_s$, the integral over $s$ can be calculated at the saddle $s=0$, and the v.e.v. can be calculated 
in the planar limit. Taking into account that 
\be
\re^{-g_s (\rho^2_N -\rho^2_{N-1})}{Z^{(0)}_{N-1} \over Z^{(0)}_N} \sim \exp\left\{  -{1\over g_s} \partial_t \left( F_0^{\IS^3}(t) -{t^3 \over 12}\right)\right\}
\ee
we find, 
\be
\label{aintrho}
A(t) ={\pi^2\over 2} +\partial_t \left( F_0^{\IS^3}(t) -{t^3 \over 12}\right)- t\int  \rd u \, \rho(u)
\log \Bigl( 4 \cosh^2 {u\over 2}
 \Bigr),\ee
where 
\be
\rho(u)={1\over \pi t} \tan^{-1} \left[ { {\sqrt{\re^t -\cosh^2 \left({u\over 2}\right)}} \over  \cosh \left({u\over 2}\right) }\right]
\ee
is the density of eigenvalues of the $\IS^3$ Chern--Simons matrix model. 

Fortunately we do not have to evaluate the integral in (\ref{aintrho}), since the result is already available in the literature. A closely related integral has been studied in 
\cite{abms,sara,jm} in the context of the $q$--deformed two--dimensional Yang--Mills theory on the two--sphere, and it calculates 
the large $N$ instanton action $\gamma(A,p)$. Its explicit expression can be found in equation (4.22) of \cite{abms}. 
In particular, for $p=2$ and $A=-4t$, (\ref{aintrho}) equals $\gamma(A,p)/4$, and using the result of \cite{abms} one finds
\be
\ba
 A(t)&= {\pi^2\over 3} -2 {\rm Li}_2(1-\re^{-t/2}) +2 {\rm Li}_2(\re^{-t/2}) +t \log (1-\re^{-t/2}) - {\rm Li}_2(\re^{-t})\\&=2 \left( {\rm Li}_2(\re^{-t/2})-  {\rm Li}_2(-\re^{-t/2})\right).
\ea
\ee
For $p=2$, the $q$-deformed Yang--Mills theory does not undergo a large $N$ phase transition, as shown in \cite{abms,sara,jm}. In our context this 
means that there are no large $N$ phase transitions when $t$ is real.

\end{document}